\documentstyle[12pt]{article}

\begin{document} 

\begin{titlepage}

\hrule 
\leftline{}
\leftline{Preprint
          \hfill   \hbox{\bf CHIBA-EP-106}}
\leftline{\hfill   \hbox{hep-th/9805153}}
\leftline{\hfill   \hbox{May 1998}}
\vskip 5pt
\hrule 
\vskip 1.0cm

\centerline{\large\bf 
Abelian Magnetic Monopole Dominance in Quark Confinement
$^*$
} 
\centerline{\large\bf  
}
\centerline{\large\bf  
}

\vskip 1cm

\centerline{{\bf 
Kei-Ichi Kondo$^{1,}{}^{\dagger}$
}}  
\vskip 1cm
\begin{description}
\item[]{\it \centerline{ 
$^1$ Department of Physics, Faculty of Science, 
Chiba University,  Chiba 263-8522, Japan}
  }
\item[]{\centerline{$^\dagger$ 
  E-mail:  kondo@cuphd.nd.chiba-u.ac.jp }
  }
\end{description}

\centerline{{\bf Abstract}} 

We prove Abelian magnetic monopole dominance in the string tension of
QCD.  Abelian and monopole dominance in low energy physics of QCD has
been confirmed for various quantities by recent Monte Carlo
simulations of lattice gauge theory. In order to prove this
dominance, we use the reformulation of continuum Yang-Mills theory
in the maximal Abelian gauge as a deformation of a topological field
theory of magnetic monopoles, which was proposed in the previous
article by the author. This reformulation provides an efficient way
for incorporating the magnetic monopole configuration as a
topological non-trivial configuration  in the functional integral.  
We derive a version of the non-Abelian Stokes theorem and use it to
estimate the expectation value of the Wilson loop.  This clearly
exhibits the role played by the magnetic monopole as an origin of
the Berry phase in the calculation of the Wilson loop in the
manifestly gauge invariant manner. We show that the string tension
derived from the diagonal (abelian) Wilson loop in the topological
field theory (studied in the previous article) converges to that of
the full non-Abelian Wilson loop in the limit of large Wilson loop.
Therefore, within the above reformulation of QCD, this result
(together with the previous result) completes the proof of quark
confinement in QCD based on the criterion of the area law of the
full non-Abelian Wilson loop.


\vskip 0.2cm
\hrule  

\vskip 0.5cm  

$^*$ To be published in Phys. Rev. D.

\end{titlepage}

\pagenumbering{arabic}

\newpage
\section{Introduction}
\setcounter{equation}{0}
\par
In a series of articles \cite{KondoI,KondoII,KondoIII}, we have
investigated quark (resp. charge) confinement in four-dimensional
non-Abelian \cite{KondoI,KondoII} (resp. Abelian \cite{KondoIII})
gauge theories.  
The main purpose of them was to clarify the mechanism of quark
(resp. charge) confinement and to give the proof of quark confinement
starting from quantum chromodynamics (QCD) (resp. quantum
electrodynamics (QED)) without introducing {\it ad hoc} assumptions.
A special gauge fixing called the maximal Abelian gauge (MAG) has
been adopted in these investigations. 
For a non-Abelian gauge group $G$, the MAG implies a partial
gauge fixing in which the coset
$G/H$ is fixed with the maximal
torus subgroup $H$ being unbroken.
The MAG is regarded as a field theoretical realization of the Abelian
projection proposed by 't Hooft \cite{tHooft81}. 
\par
In the first article \cite{KondoI}, we have proved that the QCD
vacuum is the dual  superconductor 
\footnote{
According to the recent Monte Carlo simulation, the type of dual
superconductor as the QCD vacuum is reported to be on the
border of the type II, see 
\cite{BSS98} for the defintion of the type of dual superconductor. 
This will be due to the dressing of the Abelian flux connecting
the quark and anti-quark pair by the off-diagonal gluon components,
since the Abelian dual Ginzburg-Landau theory obtained as the APEGT
is of type II (near the London limit) \cite{KondoI}. 
} in the sense that the low-energy
effective gauge theory of QCD in the MAG is given exactly by the dual
Ginzburg-Landau theory, which we called the Abelian-projected
effective gauge theory (APEGT).  This result supports  magnetic
monopole condensation as a mechanism of quark confinement. The {\it
dual superconductivity} in QCD gives the most intuitively appealing
picture of quark confinement. 
\par
In the second
article
\cite{KondoII}, we have presented a reformulation of the non-Abelian
gauge theory as a (perturbative) deformation of a topological
(quantum) field theory (T(Q)FT) which describes topological
non-trivial sector of the gauge theory. This reformulation provides
an efficient way for incorporating the magnetic monopole 
\cite{Dirac31,WY75}
configuration (which appears after Abelian projection)  as a
topological non-trivial configuration in the functional integral
of gauge theory. 
In the article
\cite{KondoII} we have defined the {\it diagonal Abelian} Wilson
loop by using the gauge field variable belonging to the maximal
torus subgroup
$H$ of  $G$.  We have proved that, in the
TFT obtained from the four-dimensional Yang-Mills (YM) theory with
a gauge group
$G$ in  the MAG, the evaluation of the  
diagonal Abelian Wilson loop is reduced to that of the equivalent
two-dimensional coset $G/H$ non-linear sigma model (NLSM).
 This equivalence is a consequence of the
Parisi-Soulous {\it dimensional reduction} of the four-dimensional
TFT in the MAG into the two-dimensional
$G/H$ coset NLSM.  This is an exact result.  This result stems from
the supersymmetry hidden in the TFT in the MAG. Moreover, we have
shown that the area law of the diagonal Wilson loop is derived by
summing up the contribution of instanton and anti-instanton
configurations in the two-dimensional NLSM.  These results lead  to
the linear confining static potential between quark and anti-quark
in the TFT sector.  For $G=SU(2)$, the equivalent model of the TFT
is given by the O(3) NLSM or
$CP^1$ model.
Thus the dimensional reduction is considered as another mechanism for
quark confinement.
\par
Similar idea can also be applied to Abelian gauge theory.  Actually,
in the third article \cite{KondoIII}, the existence of confinement
phase in the strong coupling region of QED has been shown in the
sense that the linear static potential is generated between two
fractional charges due to vortex condensation.
\par
As a background of the works \cite{KondoI,KondoII,KondoIII}, it is
necesarry to know that the Abelian and monopole dominance
\cite{EI82,SY90} in low energy physics in QCD has been confirmed
for various quantities by recent Monte Carlo simulations of lattice
gauge theory, see e.g.
\cite{review} and \cite{BOT97}.  
This is especially remarkable in the MAG. 
According to the lattice Monte Carlo simulations, the
non-Abelian string tension
$\sigma$ is nearly saturated by the Abelian part
$\sigma_{Abel}$ obtained in the MAG; Indeed, 
$\sigma_{Abel} \cong 0.92 \sigma$ for 
$G=SU(2), \beta = 2.5115$
\cite{BBMS96}.  This is called the 
{\it Abelian dominance}.  
Moreover, the Abelian part $\sigma_{Abel}$ is dominated by the
monopole contribution, $\sigma_{monopole}$ as 
$\sigma_{monopole} \cong 0.95 \sigma_{Abel}$
\cite{SNW94}.  This is called the
{\it monopole dominance}.
However, it is not clear whether the abelian and monopole dominance
on the lattice survives the continuum limit.
\par
In this article, to avoid the subtle problem of taking the continuum
limit of the lattice gauge theory, we make use of the continuum
formulation introduced in \cite{KondoII}  of the gauge theory to
study the Abelian and monopole dominance in QCD. 
Here it is important to remember that the criterion of
quark confinement should be gauge invariant, since only the gauge
invariant concept has physical meaning in gauge theories.  Indeed,
the full non-Abelian Wilson loop is gauge invariant by
construction and hence the expectation value is independent of
the gauge chosen.  Therefore the area law of the full non-Abelian
Wilson loop gives a gauge independent criterion for quark
confinement.  Consequently, the string tension obtained from the
area law is gauge-invariant and gives the gauge-independent linear
static potential between quark and anti-quark. Therefore, in the
practical calculation of the full non-Abelian Wilson loop, we can
adopt an adequate gauge so as to simplify the calculation.  It turns
out that such a simplest gauge is given by the MAG.
\par
 In this article we deal with the full non-Abelian
Wilson loop and clarify the relationship between the full non-Abelian
Wilson loop and the diagonal Wilson loop introduced and evaluated in
\cite{KondoII}.  At first glance, it seems that the area law
derived in \cite{KondoII} from the Abelian (diagonal) Wilson loop
might depend on the specific gauge fixing chosen, the MAG.   This is
not  the case, as shown in this article. 
\par
The main purpose of this article is to show that the area law of the
diagonal Wilson  loop in the TFT is sufficient to conclude the area
law of the full non-Abelian Wilson loop in the YM theory.
Actually, it turns out that  the string tension
$\sigma_{MAG}$
derived from the diagonal (abelian) Wilson loop in the TFT (studied
in the previous article \cite{KondoII}) converges to the string
tension
$\sigma$ of the full non-Abelian Wilson loop in the YM theory in the
limit of large Wilson loop $C$, that is to say, the difference
between two string tensions goes to zero in the large Wilson
loop limit,
\begin{eqnarray}
  \sigma - \sigma_{MAG} \searrow 0 \quad
  {\rm as} \quad |{\rm Area}(C)| \nearrow \infty .
  \label{mono}
\end{eqnarray}
This impiles Abelian and monopole dominance in the string tension of
QCD.  
Moreover, within the reformulation of gauge field theories given in
\cite{KondoII}, the result (\ref{mono}) completes the proof of quark
confinement in QCD based on the criterion of the area law of full
non-Abelian Wilson loop, since the area law for the diagonal Wilson
loop, i.e. 
$\sigma_{Abel}=\sigma_{MAG} \not=0$ for any value of the gauge
coupling ($g>0$) was shown using dimensional reduction and
instanton calculus in the previous article
\cite{KondoII}.
Under the MAG, we can show  without ad hoc assumptions that the dual
superconductivity and the dimensional reduction are exactly realized
in QCD, both of which lead to monopole condensations as the mechanism
for quark confinement.
\par
This article is organized as follows.
In section 2, we review the formulation of the YM theory as a
deformation of the TFT \cite{KondoII}.
In section 3, we rederive a version \cite{DP89,DP96} of the
non-Abelian Stokes theorem (NAST)
\cite{Halpern79,Bralic80,Arefeva80,Simonov89,DP89,DP96,Lunev97,HM97}
 based on the coherent state representation
\cite{Klauder79,Schulman81,KOS97,Book}.
 This clearly shows gauge invariance of the Wilson
loop and the role played by the magnetic monopole in the calculation
of the Wilson loop. The NAST clarify also the relationship between
the monopole contribution and the Berry phase
\cite{Berry84,Simon83,WZ84,Stone86,FS88,AA87,SW89}.
In section 4, the NAST is used it to estimate the expectation
value of the Wilson loop and to prove the main
statement.

\section{Yang-Mills theory as a deformation of a TFT
and dimensional reduction}
\setcounter{equation}{0}

In the previous article \cite{KondoII}, we have
presented the reformulation of the non-Abelian gauge theory as a
deformation of a topological field theory.
In this section, we summarize the essence of this reformulation for
later convenience.

\subsection{Separation of field variables}
\par
The Yang-Mills (YM) theory with a
gauge group
$G=SU(N)$ on the $D$-dimensional space-time is described by the
action $(D>2)$,
\begin{eqnarray}
 S_{QCD}^{tot}  &=& \int d^Dx ({\cal L}_{QCD}[{\cal A}_\mu,\psi]
 + {\cal L}_{GF}),
 \\
 {\cal L}_{QCD}[{\cal A}_\mu,\psi] &:=& -{1 \over 2} 
 {\rm tr}_G({\cal F}_{\mu\nu}{\cal F}_{\mu\nu})
 + \bar \psi (i \gamma^\mu {\cal D}_\mu[{\cal A}] - m) \psi ,
\end{eqnarray}
where ${\cal L}_{GF}$ is the gauge fixing term specified below and
\begin{eqnarray}
 {\cal A}_\mu(x) &=& \sum_{A=1}^{N^2-1}  {\cal A}_\mu^A(x) T^A,
 \\
 {\cal F}_{\mu\nu}(x) 
&:=& \sum_{A=1}^{N^2-1} {\cal F}_{\mu\nu}^A(x) T^A
:=   \partial_\mu {\cal A}_\nu(x) 
 -   \partial_\nu {\cal A}_\mu(x)
 - i g [{\cal A}_\mu(x), {\cal A}_\nu(x)],
  \\
  {\cal D}_\mu[{\cal A}] &:=& \partial_\mu - i g {\cal A}_\mu .
\end{eqnarray}

We adopt the following convention.
The generators $T^A(A=1, \cdots,
N^2-1)$  of the Lie algebra ${\cal G}$ of the gauge group $G=SU(N)$
are  hermitian and satisfy
$
 [T^A, T^B] = i f^{ABC} T^C ,
$
with a normalization,
$
 {\rm tr}(T^A T^B) =  {1 \over 2} \delta^{AB}.
$
Let $H=U(1)^{N-1}$ be the maximal torus group of $G$ and $T^a$ be the
generators in the Lie algebra ${\cal G}\setminus{\cal H}$ where
${\cal H}$  is the Lie algebra of $H$.
\par
In the following, we discuss only the case of $SU(2)$ explicitly,
although most of the following results can be easily extended into
$SU(N), N>2$. For $G=SU(2)$,
$T^A = (1/2) \sigma^A (A=1,2,3)$ with Pauli matrices $\sigma^A$ and
the structure constant is
$f^{ABC} = \epsilon^{ABC}$.
The indices $a, b, \cdots$ denote the off-diagonal parts of the
matrix representation.
The Cartan decomposition of the gauge field reads
\begin{eqnarray}
 {\cal A}_\mu(x) = \sum_{A=1}^3  {\cal A}_\mu^A(x) T^A
 :=  a_\mu(x)  T^3 
 + \sum_{a=1}^{2}  A_\mu^a(x) T^a .
\end{eqnarray}
\par

\par
Under the gauge transformation, the gauge field ${\cal A}_\mu(x)$
transforms as
\begin{eqnarray}
{\cal A}_\mu(x) \rightarrow  
{\cal A}_\mu^U(x) &:=& 
 U(x) {\cal A}_\mu(x) U^\dagger(x) + {i \over g}
U(x) \partial_\mu U^\dagger(x) . 
\label{GT0}
\end{eqnarray}
In order to quantize the YM theory, this gauge degrees of freedom
must be fixed by the procedure of gauge fixing.  The gauge fixing
condition is usually written as 
$F[{\cal A}]=0$. The procedure of gauge fixing must be done in
such a way that the gauge fixing condition is preserved also for the
gauge rotated field
${\cal A}_\mu^U$, i.e., $F[{\cal A}^U]=0$. 
This is guaranteed by the Faddeev-Popov (FP) ghost field.
In the BRST formalism, both the
gauge-fixing and the FP terms are automatically produced using a
functional $G_{gf}$ of the field variables as 
\begin{eqnarray}
 {\cal L}_{GF} := - i \delta_B G_{gf}[{\cal A}_\mu, {\cal C}, \bar
{\cal C},
\phi] ,
 \label{GF}
\end{eqnarray}
where ${\cal C}, \bar {\cal C}$ are ghost, anti-ghost fields and
$\phi$ is the Lagrange multiplier field for the gauge fixing
condition. Here $\delta_B$ denotes the nilpotent BRST transformation
$\delta_B$ ($\delta_B^2 \equiv 0$),
\begin{eqnarray}
   \delta_B {\cal A}_\mu(x)  &=&  {\cal D}_\mu[{\cal A}] {\cal C}(x)
   := \partial_\mu {\cal C}(x) - ig [{\cal A}_\mu(x), {\cal C}(x)],
    \nonumber\\
   \delta_B {\cal C}(x)  &=&  ig{1 \over 2} [{\cal C}(x), {\cal
C}(x)] ,
    \nonumber\\
   \delta_B \bar {\cal C}(x)  &=&   i \phi(x)  ,
    \nonumber\\
   \delta_B \phi(x)  &=&  0 ,
    \nonumber\\
   \delta_B \psi(x)  &=&   ig {\cal C}(x) \psi(x) ,
   \quad
   \delta_B \bar \psi(x) = - ig {\cal C}(x) \bar \psi(x) .
    \label{BRST0}
\end{eqnarray}
\par
The partition function of QCD is given by
\begin{eqnarray}
 Z_{QCD}[J] := \int [d{\cal A}_\mu][d{\cal C}][d\bar {\cal C}]
 [d\phi][d\psi][d\bar \psi] \exp \left\{ i S_{tot} + i S_J \right\} ,
\end{eqnarray}
with the source term,
\begin{eqnarray}
   S_J := \int d^Dx ({\rm tr}_{{\cal G}} 
   [J^\mu {\cal A}_\mu + J_c {\cal C} 
   + J_{\bar c} \bar {\cal C} + J_\phi \phi] +
\bar \eta \psi + \eta \bar \psi ).
\end{eqnarray}

\par
To reformulate the YM theory as a deformation of a topological
field theory \cite{KondoII}, we first regard the field
${\cal A}_\mu$ and
$\psi$ as the gauge transformation of the fields ${\cal V}_\mu$ and
$\Psi$,
\begin{eqnarray}
{\cal A}_\mu(x) &:=& 
 U(x) {\cal V}_\mu(x) U^\dagger(x) + \Omega_\mu(x),
 \quad \Omega_\mu(x) := {i \over g}
U(x) \partial_\mu U^\dagger(x) ,
\\
  \psi(x) &:=& U(x) \Psi(x), 
\label{GT3}
\end{eqnarray}
where ${\cal V}_\mu$ and $\Psi$ are identified with the field
variables in the perturbative sector.
Furthermore we introduce new ghost field $\gamma$, anti-ghost
field $\bar \gamma$ and the multiplier field $\beta$ which are
subject to a new BRST transformation 
$\tilde \delta_B$,
\begin{eqnarray}
   \tilde \delta_B {\cal V}_\mu(x)  &=&  {\cal D}_\mu[{\cal V}]
\gamma(x) 
   := \partial_\mu \gamma(x)  - ig [{\cal V}_\mu(x), \gamma(x) ],
    \nonumber\\
   \tilde \delta_B \gamma(x)  &=&  ig{1 \over 2} [\gamma(x) ,
\gamma(x) ] ,
    \nonumber\\
   \tilde \delta_B \bar \gamma(x)  &=&   i \beta(x)  ,
    \nonumber\\
   \tilde \delta_B \beta(x)  &=&  0 ,
    \nonumber\\
   \tilde \delta_B \Psi(x)  &=&   ig \gamma(x) \Psi(x),
   \quad
   \tilde \delta_B \bar \Psi(x) = - ig \gamma(x) \bar \Psi(x) .
    \label{BRST1}
\end{eqnarray}
Then the partition function is rewritten as
\begin{eqnarray}
 Z_{QCD}[J] &=& \int [dU][d{\cal C}][d\bar {\cal C}]
 [d\phi]
 \int [d{\cal V}_\mu][d\gamma][d\bar \gamma][d\beta]
 [d\Psi][d\bar \Psi] 
 \nonumber\\&&
\times \exp \Big\{ i \int d^Dx \Big[
 -i \delta_B G_{gf}[\Omega_\mu +  U{\cal V}_\mu U^\dagger, {\cal C}, 
 \bar {\cal C}, \phi] \Big]
 \nonumber\\&& \quad \quad \quad 
 + i \int d^D x \Big[{\cal L}_{QCD}[{\cal V}_\mu,\Psi]  
-i \tilde \delta_B \tilde G_{gf}({\cal V}_\mu, \gamma, \bar \gamma,
\beta) \Big] + i S_{J}
\Big\} ,
\label{formula}
\end{eqnarray}
where
\begin{eqnarray}
   S_J = \int d^Dx \{ {\rm tr}_{{\cal G}}[J^\mu (\Omega_\mu + 
U{\cal V}_\mu U^\dagger) + J_c {\cal C} 
   + J_{\bar c} \bar {\cal C} + J_\phi \phi] +
\bar \eta U \Psi + \eta \bar \Psi U^\dagger \} .
\end{eqnarray}

\subsection{Maximal Abelian gauge}
\par
A covariant choice for gauge fixing is the Lorentz gauge, 
\begin{eqnarray}
   F[{\cal A}] := \partial_\mu {\cal A}^\mu = 0 .
\end{eqnarray}
The most familiar choice of $G_{gf}$ is
\begin{eqnarray}
  G_{gf} = {\rm tr}_{\cal G}[\bar {\cal C}(\partial_\mu {\cal
A}^\mu + {\alpha
\over 2}\phi)] ,
\end{eqnarray}
which yields
\begin{eqnarray}
 {\cal L}_{GF} &:=& - i \delta_B G_{gf}[{\cal A}_\mu, {\cal C}, \bar
{\cal C}, \phi] 
 =  {\rm tr}_{{\cal G}}[\phi \partial_\mu {\cal A}^\mu + i \bar
{\cal C}
\partial^\mu {\cal D}_\mu[{\cal A}] {\cal C} + {\alpha \over 2}
\phi^2] .
\end{eqnarray}
The parameter $\alpha$ is called the gauge-fixing parameter.
\par
In the previous articles \cite{KondoI,KondoII}, we examined the 
maximal abelian gauge (MAG).  For $G=SU(2)$,  MAG is given by
\begin{eqnarray}
 F^{\pm}[A,a] &:=& (\partial^\mu \pm i  g a^\mu) A_\mu^{\pm} = 0,
\label{dMAG}
\end{eqnarray}
using the $(\pm, 3)$ basis,
\begin{eqnarray}
{\cal O}^{\pm} := ({\cal O}^1 \pm i{\cal O}^2)/\sqrt{2} .
\end{eqnarray}
The simplest choice of $G_{gf}$ for MAG in $(\pm, 3)$ basis is
\begin{eqnarray}
  G_{gf} = \sum_{\pm}
  \bar C^{\mp} (F^{\pm}[A,a] + {\alpha \over 2} \phi^{\pm}) ,
\label{dMAG0}
\end{eqnarray}
which is equivalently rewritten in the usual basis as
\begin{eqnarray}
  G_{gf} &=& \sum_{a=1,2}
  \bar C^{a} (F^{a}[A,a] + {\alpha \over 2} \phi^{a}) ,
\\
 F^{a}[A,a] &:=& (\partial^\mu \delta^{ab} 
   - g \epsilon^{ab3} a^\mu) A_\mu^b 
   := D^\mu{}^{ab}{}[a] A_\mu^b .
   \label{dMAG1}
\end{eqnarray}
\par
In the previous article \cite{KondoII}, we took a slightly modified
choice,
\begin{eqnarray}
G_{gf}'
= - \bar \delta_B \left( {1 \over 2} A_\mu^a A^\mu{}^a 
+ i C^a \bar C^a \right)
= - \bar \delta_B \left(
A_\mu^+ A_\mu^- + i \sum_{\pm} C^{\pm} \bar C^{\mp}
\right),
\end{eqnarray}
where $\bar \delta_B$ is the anti-BRST transformation,
\begin{eqnarray}
   \bar \delta_B {\cal A}_\mu(x)  &=&  
   {\cal D}_\mu[{\cal A}] \bar {\cal C}(x) 
   := \partial_\mu \bar {\cal C}(x) - i g [{\cal A}_\mu(x), \bar
{\cal C}(x)],
    \nonumber\\
   \bar \delta_B  {\cal C}(x)  &=&   i \bar \phi(x)  ,
    \nonumber\\
   \bar \delta_B \bar {\cal C}(x)  &=&  ig {1 \over 2} [\bar {\cal
C}(x), \bar {\cal C}(x)] ,
    \nonumber\\
   \bar \delta_B \bar \phi(x)  &=&  0 ,
    \nonumber\\
  \bar \delta_B \psi(x)  &=&    ig \bar {\cal C}(x) \psi(x) ,
\quad
  \bar \delta_B \bar \psi(x) = - ig \bar {\cal C}(x) \bar \psi(x),
\nonumber\\
\phi(x) + \bar \phi(x) &=& g [{\cal C}(x), \bar {\cal C}(x)] ,
    \label{aBRST}
\end{eqnarray}
where $\bar \phi$ is defined in the last equation.
The BRST and anti-BRST transformations have the following properties,
\begin{eqnarray}
 (\delta_B)^2 = 0, \quad (\bar \delta_B)^2 = 0, \quad
 \{ \delta_B,  \bar \delta_B \} 
 := \delta_B  \bar \delta_B + \bar \delta_B  \delta_B = 0 .
\end{eqnarray}
Our choice of gauge fixing term leads to a remarkable form for the
gauge-fixing part,
\begin{eqnarray}
{\cal L}_{GF} 
= i \delta_B \bar \delta_B \left( {1 \over 2} A_\mu^a  A^\mu{}^a
+ i C^a \bar C^a \right)
= i \delta_B \bar \delta_B \left(
A_\mu^+ A_\mu^- + i \sum_{\pm} C^{\pm} \bar C^{\mp}
\right),
\end{eqnarray}
which is invariant under the  BRST and anti-BRST transformations,
\begin{eqnarray}
 \delta_B {\cal L}_{GF} = 0 = \bar \delta_B {\cal L}_{GF} .
\end{eqnarray}

\par
The choice of $G_{gf}'$ allows the separation of the variable in such
a way
\begin{eqnarray}
{\cal L}_{GF}  &=& -i \delta_B G_{gf}'[\Omega_\mu +  U{\cal V}_\mu
U^\dagger, {\cal C},  \bar {\cal C}, \phi]
 \nonumber\\&&
 = {\cal L}_{TFT}[\Omega_\mu, {\cal C}, \bar {\cal C}, \phi]
 +  i {\cal V}_\mu^A {\cal M}_\mu^{A}[U] 
 + {i \over 2} {\cal V}_\mu^A {\cal V}_\mu^B {\cal K}^{AB}[U],
 \label{cond}
\end{eqnarray}
where we have defined,
\begin{eqnarray}
{\cal L}_{TFT} &:=& -i \delta_B G_{gf}'
[\Omega_\mu, {\cal C}, \bar {\cal C}, \phi]
= i \delta_B \bar \delta_B \left( {1 \over 2} \Omega_\mu^a
\Omega_\mu^a + i C^a \bar C^a \right)
 ,
\\
{\cal M}_\mu^A[U] &:=& \delta_B \bar \delta_B  
  [(UT^A U^\dagger)^a\Omega_\mu^a] ,
  \nonumber\\
  {\cal K}^{AB}[U] &:=& \delta_B \bar \delta_B  
  [(UT^A U^\dagger)^a(UT^B U^\dagger)^a] ,
  \label{defMK}
\end{eqnarray}
where we have used that the action of $\delta_B$ is trivial in the
perturbative sector,
\begin{eqnarray}
 \delta_B {\cal V}_\mu(x) = 0 = \bar \delta_B {\cal V}_\mu(x) .
\end{eqnarray}
The most basic BRST transformation $\delta_B$ is given by
\begin{eqnarray}
    \delta_B U(x)  =   i g {\cal C}(x) U(x) , \quad
   \bar \delta_B U(x) =  i g \bar {\cal C}(x) U(x) ,
\end{eqnarray}
which yields
\begin{eqnarray}
    \delta_B \Omega_\mu(x)  =  
   {\cal D}_\mu[\Omega]  {\cal C}(x) , \quad
   \bar \delta_B \Omega_\mu(x) =  
   {\cal D}_\mu[\Omega] \bar {\cal C}(x) .
\end{eqnarray}

\subsection{Deformation of topological field theory}
\par
The partition function of QCD is
rewritten as
\begin{eqnarray}
 Z_{QCD}[J] &:=& \int [dU][d{\cal C}][d\bar {\cal C}]
 [d\phi]
\exp \Big\{ i S_{TFT}[\Omega_\mu, {\cal C}, \bar {\cal C}, \phi] 
 \nonumber\\&&   
  +  i \int d^Dx {\rm tr}_{\cal G}[J^\mu \Omega_\mu + J_c {\cal C} 
   + J_{\bar c} \bar {\cal C} + J_\phi \phi]  
+ i W[U; J^\mu, \bar \eta, \eta]  \Big\} ,
\nonumber\\
 e^{i W[U; J^\mu, \bar \eta, \eta]}
&:=& 
 \int [d{\cal V}_\mu][d\gamma][d\bar \gamma][d\beta]
 [d\Psi][d\bar \Psi] 
 \exp \Big\{ 
 i S_{pQCD}[{\cal V}_\mu,\Psi,\gamma, \bar \gamma, \beta]
\nonumber\\&&  
 +  i \int d^Dx \Big[  {\cal V}_\mu^A {\cal J}_\mu^A
 + {i \over 2} {\cal V}_\mu^A {\cal V}_\mu^B {\cal K}^{AB}[U]
\nonumber\\&&  \quad \quad
 + {\rm tr}_{{\cal G}}(\bar \eta U \Psi + \eta \bar \Psi U^\dagger)
\Big] 
\Big\} ,
\label{deform}
\\
  S_{pQCD}[{\cal V}_\mu,\Psi,\gamma, \bar \gamma, \beta]
  &:=& \int d^Dx
\Big[ {\cal L}_{QCD}[{\cal V},\Psi]   -i \tilde \delta_B \tilde
G_{gf}({\cal V}_\mu, 
\gamma, \bar \gamma, \beta) \Big]  ,
\label{pQCD}
\\
{\cal J}_\mu^A &:=& (U^\dagger J^\mu  U)^A  + i {\cal M}_\mu^{A}[U] ,
\end{eqnarray}
where $[dU]$ is the invariant measure on the group $G$.
Here $W[U; J^\mu, \bar \eta, \eta]$ denotes the deformation from
the TFT.  When
$U \equiv 1$ and 
${\cal M}_\mu^A[U] \equiv 0 \equiv {\cal K}^{AB}[U]$,
$W[U; J^\mu, \bar \eta, \eta]$ coincides with the generating
functional of the connected correlation function in the perturbative
QCD (pQCD) with the action
$S_{pQCD}$ (topological trivial sector).
The correlation functions of
the original fundamental field
${\cal A}_\mu, \psi, \bar \psi$ is
obtained by differentiating $Z_{QCD}[J]$ with respect to the source 
$J_\mu, \bar \eta, \eta$.  
The sector written in terms of the TFT fields 
$(U, {\cal C}, \bar {\cal C},\phi)$ should be treated
non-perturbatively.  The perturbative expansion around the TFT means
the integration over the new fields 
$({\cal V}_\mu, \gamma, \bar \gamma, \beta)$ based on the
perturbative expansion in powers of the coupling constant $g$.
The deformation  $W[U; J^\mu, \bar \eta, \eta]$ should be calculated
according to the ordinary perturbation theory in the coupling
constant $g$, keeping the variable $U$ untouched.
An interpretation of this reformulation was given from the viewpoint
of the background field method \cite{Kondo98}.

\subsection{Dimensional reduction to NLSM}

For a while, we neglect the perturbative contribution $W$ and
consider only the TFT part.
Owing to the gauge choice of MAG, Parisi-Soulas dimensional reduction
occurs for the TFT.  Consequently, the D-dimensional TFT with the
action
\begin{eqnarray}
 S_{TFT}[\Omega_\mu, {\cal C}, \bar {\cal C}, \phi] 
 = \int d^{D}x \
 i \delta_B \bar \delta_B \left(
 {1 \over 2}\Omega_\mu^a(x) \Omega_\mu^a(x)
+ i C^a(x) \bar C^a(x) \right)
\label{TFT}
\end{eqnarray}
is equivalent to the
(D-2)-dimensional coset G/H non-linear sigma model (NLSM) with an
action,
\begin{eqnarray}
 S_{NLSM}[U] &=& 
 2\pi \int d^{D-2}z \ {1 \over 2} \Omega_\mu^a(z) \Omega_\mu^a(z) , 
\quad 
\Omega_\mu(z) := {i \over g} U(z) \partial_\mu U^\dagger(z)  
\nonumber\\
&=& {\beta \over 2} \int d^{D-2}z \  
  {\rm tr}_{{\cal G}\setminus{\cal H}}  
  [ \partial^\mu U(z)  \partial^\mu U^\dagger(z)] , 
  \quad \beta := {2\pi \over g^2} .
\label{NLSM}
\end{eqnarray}
Therefore, the TFT part of four-dimensional SU(2) non-Abelian gauge
theory is reduced to the two-dimensional O(3) NLSM.
Hence, the calculation of the diagonal Wilson loop for the
four-dimensional topological part is reduced to that
in the two-dimensional O(3) NLSM or equivalent $CP^1$ model. 
\par
In the previous article \cite{KondoII},
the area law decay of the expectation value of the Wilson loop in the
four-dimensional TFT (as a topological non-trivial sector of the
four-dimensional YM theory) has been deduced by summing up the
instanton and anti-instanton configurations in the two-dimensional
equivalent NLSM and the linear confining static potential between
quark and anti-quark in the TFT sector is derived. In this article, 
we show that the area law of the diagonal Wilson loop in the TFT is
sufficient to conclude the area law of the full non-Abelian Wilson
loop in the YM theory.
This completes the proof of quark confinement based on the criterion
of the Wilson loop in the four-dimensional YM theory.

\section{Non-Abelian Stokes theorem}
\setcounter{equation}{0}

The Wilson loop operator is defined as a path ordered product of an
exponent along a closed loop $C$.  
In the Abelian case, due to the
ordinary Stokes theorem, it is rewritten as a surface integral on the
surface
$S$ whose boundary is given by $C$.  
In contrast to the ordinary Stokes theorem, there may be many
possibilities for the non-Abelian Stokes theorem (NAST)
\cite{Halpern79,Bralic80,Arefeva80,Simonov89,DP89,DP96,Lunev97,HM97}.  
In this article we treat a version of the NAST derived by Diakonov
and Petrov \cite{DP89,DP96}.  This version of NAST is able to remove
path ordering from the expression of the non-Abelian Wilson loop.
Instead, we must perform the functional integration.
First of all, we rederive the NAST for the gauge group
$G=SU(2)$ using the path integral formalism in the spin coherent
state representation.
Moreover, we clarify the relationship between the induced
magnetic monopole and the Berry phase which appear in the
NAST.
Second, we give the general NAST for any compact Lie group $G$. The 
NAST is manifestly gauge invariant as in the
Wilson loop. In the next section, we make use it to prove the
abelian monopole dominance in the string tension of QCD.

\subsection{Path integral in the coherent state representation}

We consider the  formal expression,
\begin{eqnarray}
 Z[t,0] := {\rm tr} {\cal P}_t \exp \left[ - i \int_0^t d\tau
 {\cal H}(\tau) \right] ,
\end{eqnarray}
where ${\cal P}_\tau$ is the $t$-ordering (or path-ordering)
operator and the "Hamiltonian" ${\cal H}$ is specified later. 
We can make it well defined by taking the limit of discretization,
\begin{eqnarray}
 Z[t,0] = \lim_{N \rightarrow \infty, \epsilon \rightarrow 0}    
{\rm tr} {\cal P}_\tau
\prod_{n=0}^{N-1}  [1- i \epsilon {\cal H}(\tau_n) ],
\label{Z2}
\end{eqnarray}
where   
$\epsilon=t/N$ is the timestep and 
$\tau_n=n \epsilon$ is the discrete time. 
The limit is taken keeping $N\epsilon=t$ constant.
For a given Hamiltonian in the representation $J$, we would like to
obtain a path integral representation of the partition function
for a spin system. 
\par
We make use of the spin coherent state to write the path integral
representation of $Z[t,0]$.
Consider the group SU(2) and an irreducible representation
characterized by highest spin $J$. Let 
$|0 \rangle$ denote the maximally polarized state
$|0 \rangle=|J,J \rangle$ which is the highest weight state of a
spin-$J$ representation 
$\{|J,M \rangle \}$ of $SU(2)$ where $M$ labels the eigenvalue of
$J_3$,
\begin{eqnarray}
 {\bf J}^2 |J,M \rangle &=& J(J+1) |J,M \rangle ,
 \quad (-J \le M \le J) ,
 \nonumber\\
 J_3 |J,M \rangle &=& M |J,M \rangle  .
\end{eqnarray}
The state $|J,M \rangle$ is an eigenvector of both the diagonal
generator $J_3$ (Cartan subalgebra) and the quadratic Casimir
invariant ${\bf J}^2$. Spin coherent sates are a family of spin state
$\{ |{\bf n} \rangle \}$ which is obtained by applying
 the rotation operator $R$ to the maximally
polarized state $|0 \rangle$,
\begin{eqnarray}
 |{\bf n} \rangle := R(\chi, \theta, \varphi) |J, J \rangle
 = e^{iJ^3 \varphi} e^{i J^2 \theta} e^{iJ^3 \chi} |J, J \rangle ,
 \label{rep}
\end{eqnarray}
where $J^A(A=1,2,3)$ are three generators of SU(2) and
$(\chi,\varphi,\theta)$ are Euler angles and the unit vector
${\bf n}$ parameterizes the spin coherent state.  We have the freedom
to define
$\chi$ arbitrary.  This is a U(1) gauge freedom.  We can eliminate
it by fixing $\chi$.  This is the gauge fixing for the residual
gauge group $H=U(1)$.  
The states are in one-to-one correspondence with the (right) coset
$SU(2)/U(1)$ where $U(1)$ is generated by $J_3$ (rotation about the
$z$ axis).
In the language of differential geometry, the coherent states
form a Hermitian line bundle associated with the Hopf, or monopole,
principal bundle. 
\par
The diagonal matrix element of the generators reads
\begin{eqnarray}
 \langle {\bf n} | J^A |{\bf n} \rangle  =  J n^A ,
\label{formula1}
\end{eqnarray}
where
\begin{eqnarray}
  {\bf n}(x)  = (n^1(x),n^2(x),n^3(x)) 
  = (\sin \theta(x) \cos \varphi(x),
\sin \theta(x) \sin \varphi(x), \cos \theta(x) ).
\label{ndef}
\end{eqnarray}
It is known \cite{Klauder79,Book} that the coherent sates are not
orthogonal.  The overlap, i.e. the inner product of any two coherent
states is evaluated as
\begin{eqnarray}
 \langle {\bf n}  |{\bf n}' \rangle &=&
 \left( {1 + {\bf n} \cdot {\bf n}' \over 2} \right)^J
e^{-iJ\Phi({\bf n},{\bf n}')},
 \\
 \Phi({\bf n},{\bf n}') &:=& 2 \arctan \left\{ 
 {\cos [{1 \over 2}(\theta+\theta')] \over
 \cos [{1 \over 2}(\theta-\theta')]}
 \tan \left( {\varphi - \varphi' \over 2} \right) \right\} +
\chi - \chi' ,
\label{formula2}
\end{eqnarray}
where $\chi, \chi'$ depend on the gauge fixing. 

\par
  The coherent states span the space of states of spin $J$.  The
measure of integration over the group parameters is defined by
\begin{eqnarray}
  d\mu({\bf n}) 
 := {2J+1 \over 4\pi}  \delta({\bf n}\cdot {\bf n}-1) d^3 {\bf n}
 = {2J+1 \over 4\pi} \sin \theta d\theta d\varphi .
\end{eqnarray}
This is a Haar measure of the coset $SU(2)/U(1)$, in other words,
it is the area element on the two-sphere $S^2$.
The state $ |{\bf n} \rangle$ can be expanded in a complete basis of
the spin-$J$ irreducible representation 
$\{|J,M \rangle \}$.
The coefficients of the expansion are the representation matrix,
\begin{eqnarray}
  | {\bf n} \rangle = \sum_{M=-J}^{+J} |J,M \rangle 
 D^{(J)}_{MJ}({\bf n}) .
\end{eqnarray}
The resolution of unity is given by
\begin{eqnarray}
  \int d\mu({\bf n})
 |{\bf n} \rangle \langle {\bf n}| 
 = \sum_{M=-J}^{+J} |J,M \rangle \langle J,M | = I ,
 \label{formula3}
\end{eqnarray}
where $I$ is an identity operator.
Hence the coherent state $ |{\bf n} \rangle$ forms the complete set,
although it is not orthogonal. Thus the coherent states form an
overcomplete basis.
\par
In particular, for $J={1 \over 2}$, an element $U(x) \in SU(2)$ is
written as follows by introducing three local field variables 
$(\theta(x), \varphi(x), \chi(x))$ corresponding to the
Euler angles, 
\begin{eqnarray}
 R(\varphi, \theta, \chi) &=& U(x) 
 = e^{i \varphi(x) \sigma_3/2} 
   e^{i \theta(x) \sigma_2/2} 
   e^{i \chi(x) \sigma_3/2}  
   \nonumber\\
 &=& \pmatrix{
 e^{{i \over 2}(\varphi(x)+\chi(x))} \cos {\theta(x) \over 2} &
 -e^{{i \over 2}(\varphi(x)-\chi(x))} \sin {\theta(x) \over 2} \cr
 e^{-{i \over 2}(\varphi(x)-\chi(x))} \sin {\theta(x) \over 2} &
 e^{-{i \over 2}(\varphi(x)+\chi(x))} \cos {\theta(x) \over 2} } ,
\nonumber\\&&
 \theta \in [0,\pi], 
\varphi \in [0,2\pi], \chi \in [0, 2\pi] ,
\label{Euler}
\end{eqnarray}
and (\ref{rep}) reads
\begin{eqnarray}
 |{\bf n} \rangle = R(\chi, \theta, \varphi) \pmatrix{1 \cr 0}
 = e^{{i \over 2}\chi(x)} \pmatrix{ 
 e^{{i \over 2}\varphi(x)} \cos {\theta(x) \over 2} \cr 
  e^{-{i \over 2}\varphi(x)} \sin {\theta(x) \over 2}  } .
\end{eqnarray}
By making use of the explicit representation, we can make sure that
the formulae (\ref{formula1}), (\ref{formula2}) and (\ref{formula3})
hold for $J=1/2$.

\par
Inserting $N$ resolutions of unit (\ref{formula3}) between the
factors in (\ref{Z2}), we obtain 
\begin{eqnarray}
 Z[t,0] = \lim_{N \rightarrow \infty, \epsilon \rightarrow 0 }
   \prod_{n=1}^{N} 
 d\mu({\bf n}(\tau_n)) 
\prod_{\tau=\epsilon}^{t}  
\langle {\bf n}(\tau) | {\bf n}(\tau-\epsilon) \rangle 
 [1 - i\epsilon H(\tau) ],
 \label{Z3}
\end{eqnarray}
where the "classical" Hamiltonian is defined by
\begin{eqnarray}
 H(\tau) := { 
 \langle {\bf n}(\tau) | {\cal H}(\tau) |{\bf n}(\tau-\epsilon)
\rangle \over 
\langle {\bf n}(\tau) | {\bf n}(\tau-\epsilon) \rangle} ,
\end{eqnarray}
and the periodic boundary condition is adopted,
\begin{eqnarray}
 {\bf n}(t) = {\bf n}(0) .
\end{eqnarray}
In the limit $N \rightarrow \infty$, we replace the differences
by the corresponding derivatives,
\begin{eqnarray}
   {\bf n}(\tau+\epsilon) - {\bf n}(\tau) 
  \rightarrow \epsilon \dot {\bf n}(\tau)  + O(\epsilon^2) .
  \label{diff}
\end{eqnarray}
For more rigorous treatment, see \cite{Klauder79,Schulman81}.
\par
Using (\ref{formula2}), the overlap between coherent states at nearby
steps to leading order in $\epsilon$ reads
\begin{eqnarray}
&& \prod_{\tau=\epsilon}^{t}  
\langle {\bf n}(\tau+\epsilon) | {\bf n}(\tau) \rangle 
\nonumber\\
&=& \exp \left\{ - i J \sum_{\tau=\epsilon}^{t}  \Phi({\bf
n}(\tau+\epsilon), {\bf n}(\tau)) 
+ J \sum_{\tau=\epsilon}^{t} \ln \left[ {1+ {\bf n}(\tau) \cdot {\bf
n}(\tau-\epsilon) \over 2} \right]  \right\} .
\end{eqnarray}
Making use of (\ref{formula2}), we obtain
\begin{eqnarray}
 \Phi({\bf n}(\tau+\epsilon), {\bf n}(\tau)) 
 \cong  \epsilon  
 [ \dot \varphi(\tau) \cos \theta(\tau) + \dot \chi(\tau)] , 
\end{eqnarray}
whereas (\ref{diff}) leads to
\begin{eqnarray}
 \ln \left[ {1+ {\bf n}(\tau) \cdot {\bf
n}(\tau-\epsilon) \over 2} \right] 
 \cong  \ln \left[ 1 - {\epsilon^2 \over 4} (\partial_\tau {\bf n})^2
\right] \cong  - {\epsilon^2 \over 4} (\partial_\tau {\bf n})^2 .
\end{eqnarray}
Within the same approximation, the classical Hamiltonian can be
evaluated at equal times,
\begin{eqnarray}
 H(\tau)  \rightarrow
 \langle {\bf n}(\tau) | {\cal H}(\tau) |{\bf n}(\tau)
\rangle  + O(\epsilon) .
\end{eqnarray}
By exponentiating the Hamiltonian and discarding higher-order terms
in $\epsilon$, the formal continuum limit of (\ref{Z3}) is obtained,
\begin{eqnarray}
  Z[t,0] &=& \int [d\mu_C({\bf n})] \exp ( i S[{\bf n}]),
  \\
  S[{\bf n}] &:=& - \int_0^t d\tau H[{\bf n}] 
  - \gamma (t) 
    + {J \over 4} \epsilon \int_0^t d\tau  (\partial_\tau {\bf n})^2 
,
    \\
  H[{\bf n}] &:=& \langle {\bf n}(\tau) | {\cal H}(\tau) |{\bf
n}(\tau)  \rangle  ,
  \\
  \gamma (t) &:=& J \int_0^t d\tau
  [\dot \varphi(\tau) \cos \theta(\tau) + \dot \chi (\tau)] ,
\end{eqnarray}
where
\begin{eqnarray}
 [d\mu_C({\bf n})] &:=& \lim_{N \rightarrow \infty, \epsilon
\rightarrow 0}  
  \prod_{n=1}^{N} d\mu({\bf n}(\tau_n)) .
\end{eqnarray}
\par
The first term, the Hamiltonian, is specified below.
Though the last term in $S[{\bf n}]$ vanishes in the continuum limit
$\epsilon \rightarrow 0$, it plays the role of a regularization. 
Without it, the 'action' $S[{\bf n}]$ has no 'kinetic term' for the
field 
${\bf n}$.
\par
The second term $\gamma (t)$ depends on the trajectory of ${\bf
n}(\tau)$ on the sphere and not on its explicit time dependence.  It
is geometric.
The phase $\gamma (t)$ is called the {\it Berry phase} or
{\it geometric phase} of the spin history \cite{Berry84,Simon83}.  
The Berry phase measures the area enclosed by the path
${\bf n}(\tau)$  on the unit sphere.  
The area increment is a spherical triangle with vertices at 
${\bf n}(\tau), {\bf n}(\tau+ \epsilon)$ and $(0,0,1)$  whose area
is given by
\begin{eqnarray}
  \omega := [1-\cos \theta(\tau)] d\varphi(\tau) .
\end{eqnarray}
Hence the total area enclosed by the closed orbit is equal to
\begin{eqnarray}
 \oint_\Gamma \omega :=
 \int_0^t d\tau [1-\cos \theta(\tau)] \dot \varphi(\tau). 
\end{eqnarray}
The Berry phase 
\begin{eqnarray}
  \gamma(t) = J \Omega = J \oint_\Gamma \omega 
  = 4 \pi J Q 
\end{eqnarray}
is expressed in a gauge invariant form.

We can introduce a vector potential 
\begin{eqnarray}
 \Omega :=
 \int_0^t d\tau {\bf A}(\tau) \cdot {d \over d\tau}{\bf n}(\tau) . 
\end{eqnarray}
producing the a unit magnetic monopole whose line integral over the
orbit ${\bf n}(\tau)$ is equal to the solid angle $\Omega$ subtended
by that orbit.
For example, in the domain 
 \begin{eqnarray}
  U_S &:=&  S^2 - {\rm South~Pole} 
  = \{ (\theta, \varphi) \in S^2; \theta \not= \pi \} ,
  \\
   U_N &:=&     S^2 - {\rm North~Pole} 
   = \{ (\theta, \varphi) \in S^2; \theta \not= 0 \} ,
   \\
   U_M &:=&    S^2 - {\rm Meridian}
   = \{ (\theta, \varphi) \in S^2; 
   \theta \not= 0, \pi ,  \varphi \not= 0 \} ,
\end{eqnarray}
the vector potential is respectively given by
\begin{eqnarray}
  {\bf A}_S &:=&
  - {1 - \cos \theta \over r \sin \theta} \hat \varphi 
  = - \left( - {y \over r(r+z)}, {x \over r(r+z)}, 0 \right) ,
  \nonumber\\
    {\bf A}_N &:=& {1 + \cos \theta \over r \sin \theta} \hat
\varphi 
  = \left( - {y \over r(r-z)}, {x \over r(r-z)}, 0 \right) ,
  \nonumber\\
   {\bf A}_M &:=& { \cos \theta \over r \sin \theta} \hat \varphi 
  = \left( - {yz \over r(r^2-z^2)}, {xz \over r(r^2-z^2)}, 0 \right)
,
\label{vecp}
\end{eqnarray}
where $\hat \varphi$ is a unit vector in the direction of $\varphi$.
The corresponding connection one-form $\omega$ is given (for a
choice of $\chi$) by
\begin{eqnarray}
  \omega_S &:=&  J (- \cos \theta+1) d \varphi 
  = 2J \sin^2{\theta \over 2} d \varphi, 
  \quad (\chi = - \varphi), 
  \nonumber\\
  \omega_N &:=& J (- \cos \theta-1) d \varphi 
  = - 2J \cos^2 {\theta \over 2} d \varphi ,
  \quad (\chi =  + \varphi), 
  \nonumber\\
  \omega_M &:=&  - J \cos \theta d \varphi .
  \quad (\chi = 0) . 
  \label{conn}
\end{eqnarray}
Note that $\omega_N$ and $\omega_S$ are interrelated by the gauge
transformation,
\begin{eqnarray}
  \omega_S  =  \omega_N  + 2J d \varphi .
  \label{gc}
\end{eqnarray}
The connection one-form is related to the curvature two-form by the
ordinary Stokes theorem,
\begin{eqnarray}
  \oint_{\Gamma} \omega  =  \int_S d\omega,  \quad
  \Gamma=\partial S .
\end{eqnarray}
The curvature two-form  $d\omega$ does not depend on the
choice of the connection one-form (\ref{conn}), since
\begin{eqnarray}
  d\omega = J \sin \theta d\theta \wedge d\varphi  
\end{eqnarray}
The Berry phase measures  the flux of magnetic
monopole through the area $S$ of $S^2$ bounded by the trajectory
$\Gamma$ of
${\bf n}(t)$.
\par
Perform the contour integral along the $\theta$= constant line for
(\ref{gc}),
\begin{eqnarray}
  \oint_C \omega_S  =  \oint_C \omega_N  + 2J \oint_C d \varphi 
  = \oint_C \omega_N  + 4\pi J  .
  \label{gc2}
\end{eqnarray}
This implies
\begin{eqnarray}
  e^{i \oint_C \omega_S}
  = e^{i \oint_C \omega_N} e^{i4\pi J} ,
  \label{gc3}
\end{eqnarray}
and $e^{i4\pi J}=1$, i.e., $4\pi J=2\pi n$.
Thus the quantization of the spin $J={n \over 2}$ is obtained as a
topological invariant.
Incidentally, the connection one-form $\omega_S, \omega_N$
is written using the unit vector as
\begin{eqnarray}
 \omega(x)   = J {n^1(x) dn^2(x) - n^2(x) dn^1(x) \over 1 \pm
n_3(x)} .
\end{eqnarray}

\subsection{Non-Abelian Stokes theorem for $G=SU(2)$}
\par
Now we apply the above result to evaluate the Wilson loop
operator.
We consider the Hamiltonian
\begin{eqnarray}
  {\cal H}(t) 
  =  {\cal A}(t) :=  {\cal A}_\mu(x) {dx^\mu \over dt}
  =  {\cal A}^A(t) T^A
  =  {\cal A}_\mu^A(x) T^A {dx^\mu \over dt} ,
\end{eqnarray}
where ${\cal A}(t)$ is the tangent component of the YM field along
the loop (see the next subsection for a more precise definition).
Using (\ref{formula1}), we obtain
\begin{eqnarray}
 H[{\bf n}]  = i J {\cal A}^A(t) n^A(t) 
 = J {\rm tr}[\sigma_3 U {\cal A} U^\dagger] ,
\end{eqnarray}
and
\begin{eqnarray}
 \int_0^t d\tau H[{\bf n}]  
 =  J \int_0^t d\tau  \ {\rm tr}[\sigma_3 U {\cal A} U^\dagger] . 
\end{eqnarray}
where $\sigma_3$ is the third Pauli matrix and we have used  the
adjoint orbit representation for ${\bf n}$, 
\begin{eqnarray}
 n^A(x) T^A = U^\dagger(x) T^3 U(x)  . 
 \label{aorep}
\end{eqnarray}
Using (\ref{Euler}), we can see that
the unit vector ${\bf n}(x)$ defined by (\ref{aorep}) is equal to 
(\ref{ndef}).
\par
On the other hand, using (\ref{Euler}) the Berry phase is
rewritten as
\begin{eqnarray}
    \gamma(t) = J \int_0^t d\tau {\rm tr}
  \left( \sigma_3 i U {d \over dt}U^\dagger \right) ,
\end{eqnarray}
where the functional $\gamma (t)$ denotes the
phase acquired by a spin that aligns with an adiabatically rotating
external field $\vec {\cal A}$ which is parallel to 
${\bf n}(\tau)$.  
 Finally we have shown 
\begin{eqnarray}
  Z[t,0]  &=&   {\rm tr} {\cal P}_C \exp \left[ - i \oint_C dx^\mu
 {\cal A}_\mu(x)   \right] 
 = \int [d\mu({\bf n})] \exp ( i S[{\bf n}]),
  \\
  S[{\bf n}] &:=&  J \int_0^t d\tau  \
   {\rm tr} \left\{ \sigma_3 \left( U {\cal A} U^\dagger  
 +  i U {d \over d\tau}U^\dagger \right) \right\} .
\end{eqnarray}
\par
For the gauge group G=SU(2), we have obtained the non-Abelian Wilson
loop in the path integral representation,
\begin{eqnarray}
 && W^C [{\cal A}] :=  {\rm tr} \left[ {\cal P} 
 \exp \left( i \oint_C {\cal A}_\mu^A(x) T^A dx^\mu
 \right) \right] 
 \nonumber\\ 
 &=& 
 \int  [d\mu_C({\bf n})]  
 \exp \left( i J \oint_C dx^\mu \ {\rm tr}
 \left\{  \sigma_3 \left[ 
  U {\cal A}_\mu(x)U^\dagger 
  + {i \over g} U \partial_\mu U^\dagger  \right]
 \right\} \right) 
 \nonumber\\ 
 &=&  \int  [d\mu_C({\bf n})]  
 \exp \left( i J \oint dt \ {\rm tr} \left\{ \sigma_3
\left[  U {\cal A}(t) U^\dagger 
 + {i \over g} U {d \over dt} U^\dagger \right] \right\}  \right) ,
 \label{NAST}
\end{eqnarray}
where $J$ is the spin of the
representation of the Wilson loop considered.
This is a special case of the NAST of Diakonov and
Petrov which will be explained in the next section.

\subsection{Non-Abelian Stokes theorem in the general case}

\par
We give the results of Diakonov and Petrov \cite{DP89,DP96} in the
most general form in the following.

{\it Definition}:
Let $C$ be a given curve 
$x_\mu=x_\mu(t)$ parameterized by $t$ where
the values of the parameter $t_1,t_2$ correspond to the end points
of the curve.  
We define the path-ordered exponent (POE) of the YM
field ${\cal A}_\mu(x) := {\cal A}_\mu^A(x) T^A$ by
\begin{eqnarray}
 W^C_{\alpha,\beta}(t_2,t_1) := \left[ {\cal P} 
 \exp \left( i \int_{x(t_1)}^{x(t_2)} {\cal A}_\mu^A(x) T^A dx^\mu
 \right) \right]_{\alpha,\beta} .
 \label{POE}
\end{eqnarray}
Introducing the tangent component of the YM field along the loop,
\begin{eqnarray}
  {\cal A}(t) := {\cal A}_\mu(x) {dx^\mu \over dt}
  = {\cal A}_\mu^A(x) T^A {dx^\mu \over dt} ,
\end{eqnarray}
we can write the POE as
\begin{eqnarray}
 W^C_{\alpha,\beta}(t_2,t_1) := \left[ {\cal P} 
 \exp \left( i \int_{t_1}^{t_2}  {\cal A}(t) dt
 \right) \right]_{\alpha,\beta} .
 \label{POE2}
\end{eqnarray}
The POE (\ref{POE}) is defined by the power-series expansion,
\begin{eqnarray}
 W^C_{\alpha,\beta}(t_2,t_1) 
 = \sum_{n=0}^{\infty} \int d\tau_1 \cdots \int d\tau_n  
 [i {\cal A}(\tau_1) \cdots i {\cal A}(\tau_n)]_{\alpha,\beta} ,
 \label{POE3}
\end{eqnarray}
where 
$t_2 \ge \tau_1 \ge \cdots \ge \tau_n \ge t_1$. 

{\it Theorem}\cite{DP89,DP96}: Consider the non-Abelian group $G$
and the maximal torus group $H$ of $G$.  Define
$T^A$ to be the generators of the representation $J$: 
$T^A T^A = J(J+1)$.  
Let ${\cal H}_i (i=1,\cdots, r)$ be the generators of the Cartan
subalgebra of the Lie algebra
${\cal G}$ of $G$ and the
$r$-dimensional vector ${\bf m}$ be the highest weight of the
representation $J$ with  $r$ being the rank of the gauge group $G$. 
Then the POE is written in the path integral form,
\begin{eqnarray}
 && W^C_{\alpha,\beta}(t_2,t_1) 
 \nonumber\\ 
 &=& \int dU_1 \int dU_2 \sum_{J',M'} (2J'+1) 
 D^{(J')}_{\alpha M'}(U_2^\dagger) D^{(J')}_{M'\beta}(U_1)
 \nonumber\\&& \times
 \int_{U(t_1)=U_1}^{U(t_2)=U_2} [dU(t)] 
 \exp \left( i J \int_{x(t_1)}^{x(t_2)} dx^\mu \ {\rm tr}
 \left\{ m_i {\cal H}_i \left[ 
  U {\cal A}_\mu(x)U^\dagger 
  + {i \over g} U \partial_\mu U^\dagger  \right]
 \right\} \right) 
 \nonumber\\ 
 &=& \int dU_1 \int dU_2 \sum_{J',M'} (2J'+1) 
 D^{(J')}_{\alpha M'}(U_2^\dagger) D^{(J')}_{M'\beta}(U_1)
 \nonumber\\&& \times
 \int_{U(t_1)=U_1}^{U(t_2)=U_2} [dU(t)] 
 \exp \left( i J \int_{t_1}^{t_2} dt \ {\rm tr}
 \left\{ m_i {\cal H}_i 
\left[  U {\cal A}(t) U^\dagger 
 + {i \over g} U {d \over dt} U^\dagger \right]
\right\}  \right) .
\end{eqnarray}
Here $dU$ is the invariant Haar measure on $G/H$ 
and $D^{T}_{MM'}(U)$ are the Wigner $D$-function which
expresses finite rotation in the representation $J$,  
\begin{eqnarray}
  R(U) |J, M \rangle = \sum_{M'=-J}^{+J} |J, M' \rangle
  D^{(J)}_{M'M}(U) ,
  \quad
   D^{(J)}_{MM'}(U) := \langle J, M | R(U) |J, M' \rangle .
\end{eqnarray}
In particular,
in the spinor representation 
$D^{1/2}_{MM'}(U)=U_{MM'}$.

\par

According to the above theorem, the POE is written as a functional
integral over all gauge transformations $U(t)$ of the given
potential ${\cal A}(t)$, projected into matrix representation 
$\alpha, \beta$.
From the above theorem, a version of NAST is
given as follows.
The Wilson loop, i.e. the trace of POE
along a closed loop
$C$ has the form,
\begin{eqnarray}
 && W^C [{\cal A}] :=  {\rm tr} \left[ {\cal P} 
 \exp \left( i \oint_C {\cal A}_\mu^A(x) T^A dx^\mu
 \right) \right] 
 \nonumber\\ 
 &=& 
 \int [dU(t)] 
 \exp \left( i J \oint_C dx^\mu \ {\rm tr}
 \left\{  m_i {\cal H}_i \left[ 
  U {\cal A}_\mu(x)U^\dagger
  + {i \over g} U \partial_\mu U^\dagger  \right]
 \right\} \right) 
 \nonumber\\ 
 &=&  \int  [dU(t)] 
 \exp \left( i J \oint dt \ {\rm tr} \left\{  m_i {\cal H}_i 
\left[  U {\cal A}(t) U^\dagger 
 + {i \over g} U {d \over dt} U^\dagger \right]
\right\}  \right) .
\end{eqnarray}
For $G=SU(2)$, this reduces to (\ref{NAST}).
The formula is manifestly gauge invariant, as is the Wilson loop
itself.

\section{Abelian and monopole dominance}
\setcounter{equation}{0}

\par
Now we show that the abelian and monopole dominance is deduced from
the NAST in the formulation \cite{KondoII} of YM theory as a
deformation of the MAG TFT. Making use of the NAST, we will clarify
the meaning of Abelian dominance and monopole dominance in
low-energy physics in QCD. 
\par
The full non-Abelian Wilson loop is defined as the path-ordered
exponent. In the version of the NAST derived in the previous
section, the path ordering has been removed from the expression. 
Instead, we must average over the Haar measure on $G/H$.
The removal of path ordering is very welcome, since it is rather
difficult to treat the path ordering.  As a result, there appears
the field tensor introduced by 'tHooft and Polyakov in connection
with magnetic monopoles.
This indicates an intimate connection between the magnetic monopole
and quark confinement.
In what follows, explicit calculations are performed only for
$G=SU(2)$.  However, the generalization to $G=SU(N)$ is
straightforward.

\subsection{Magnetic monopole in YM theory}
\par
The non-perturbative study of YM theory in the MAG goes as follows,
see \cite{KondoI} for more details. First of all, gauge field
configurations are constructed by performing the local gauge
transformation (\ref{GT0}) such that the gauge rotated field
${\cal A}_\mu^U(x)$  minimize the functional ${\cal R}[{\cal A}^U]$ 
where
\begin{eqnarray}
  {\cal R}[{\cal A}] := \int d^Dx  \ {\rm tr}_{{\cal G}\setminus{\cal
H}}[ {1
\over 2} {\cal A}_\mu(x) {\cal A}_\mu(x)] .
  \label{MAG2}
\end{eqnarray}
Here the trace is taken over the Lie algebra ${\cal G}\setminus{\cal
H}$.
In the differential form, this implies that
${\cal A}_\mu^U(x)$ satisfies the the gauge-fixing condition
(\ref{dMAG}). Next, the Abelian (or diagonal) field $a_\mu^U$ and
its field strength $f_{\rho\sigma}$ are extracted from the
non-Abelian gauge field according to
\begin{eqnarray}
 a_\mu^U(x) &:=& {\rm tr}[T^3 {\cal A}_\mu^U(x)] ,
\\
 f_{\mu\nu}^U(x) &:=& \partial_\mu a_\nu^U(x) - \partial_\nu
a_\mu^U(x) .
\label{Abef}
\end{eqnarray}
The  magnetic current $k_\mu$ is defined from the
diagonal part by
\begin{eqnarray}
  k_\mu(x)  =   \partial_\nu \tilde f_{\mu\nu}^U(x) ,
\quad
\tilde f_{\mu\nu}(x) := {1 \over 2}
\epsilon_{\mu\nu\rho\sigma} f_{\rho\sigma}(x) .
\end{eqnarray}
\par
The magnetic charge is calculated from the magnetic
current as
\begin{eqnarray}
 g_m(V^{(3)}) = \int_{V^{(3)}} d^3 \sigma_\mu k_\mu 
 =   \int_{V^{(3)}} d^3 \sigma_\mu
 \partial_\nu \tilde f_{\mu\nu}^\Omega
=   \int_{S^{(2)}=\partial V^{(3)}} d^2
\sigma_{\mu\nu} \tilde f_{\mu\nu}^\Omega .
\end{eqnarray}
In the usual Abelian gauge theory (i.e., Maxwell theory), the
magnetic monopole current vanishes identically due to the Bianchi
identity,
\begin{eqnarray}
 \epsilon_{\mu\nu\rho\sigma}  \partial^\mu f^{\rho\sigma}(x) \equiv
0 ,
\quad
 f_{\mu\nu}(x) := \partial_\mu a_\nu(x) - \partial_\nu a_\mu(x) ,
\end{eqnarray}
as long as the field variable $a_\mu(x)$ is non-singular.
In other words, in order to obtain a non-trivial magnetic current in
Abelian gauge theory, we need to introduce the singularity into the
Abelian gauge theory.  This fact is well known from the study of
Dirac magnetic monopole.
In the non-Abelian gauge theory, the
singularity is produced by  partially fixing the gauge
$G/H$ and leaving the Abelian subgroup $H$ of the original
non-Abelian gauge group $G$ unbroken. The partial gauge fixing leads
to the singularity which is sufficient to generate the magnetic
monopole. This is an idea of Abelian projection by 't Hooft
\cite{tHooft81}. 
The MAG leaves the maximal torus group $H=U(1)^{N-1}$
unbroken.  This is why the YM theory can have magnetic monopoles
even in the absence of the Higgs scalar field.  
It is well known that the YM theory in Euclidean space has instanton
solutions, although the pure YM theory does not have any non-trivial
classical (stable) soliton solution in four-dimensional Minkowski
spacetime. 
It is still in dispute whether the instanton configuration alone can
produce sufficient string tension for the quark confinement.
The relationship between the magnetic monopole and instanton has
been discussed in \cite{KondoI,KondoII}, see also references cited
there.
\par
Substituting (\ref{GT0}) into (\ref{Abef}), we have
\begin{eqnarray}
 a_\mu^U(x) &:=& {\rm tr}[T^3 {\cal A}_\mu^U(x)] 
 =  n^A(x) {\cal A}_\mu^A(x)  + a_\mu^\Omega(x) ,
 \label{decomp1}
\end{eqnarray}
where we have used (\ref{aorep}) and defined
\begin{eqnarray}
 a_\mu^\Omega(x) := \Omega_\mu^3(x) :={\rm tr}[T^3 \Omega_\mu(x)] ,
 \quad
 \Omega_\mu(x) :=  {i \over g} U(x) \partial_\mu U^\dagger(x) .
\end{eqnarray}
\par
Note that (\ref{decomp1}) has the same form as the argument of the
exponent in the NAST (\ref{NAST}).
Therefore the NAST (\ref{NAST}) for the Wilson loop is rewritten as
\begin{eqnarray}
  W^C[{\cal A}] = \int [d\mu_C({\bf n})] \exp \left( iJ \oint_C 
  dx^\mu a_\mu^U (x) \right) ,
  \label{fWl}
\end{eqnarray}
and the expectation value of the the Wilson loop is given by
\begin{eqnarray}
  \langle W^C[{\cal A}] \rangle_{YM} = \int [d\mu_C({\bf n})]
\left\langle
\exp
\left( iJ \oint_C   dx^\mu a_\mu^U (x) \right) \right\rangle_{YM}.
\label{fWle}
\end{eqnarray}
\par
In the previous article \cite{KondoII} we have calculated the 
expectation value 
\begin{eqnarray}
  \langle W^C[a^\Omega] \rangle_{YM} = 
\left\langle
\exp \left( iJ \oint_C   dx^\mu a_\mu^\Omega (x) \right)
\right\rangle_{YM} 
\end{eqnarray}
of the diagonal Wilson loop,
\begin{eqnarray}
  W^C[a^\Omega] =  \exp \left( iJ \oint_C 
  dx^\mu a_\mu^\Omega (x) \right),
  \quad a_\mu^\Omega(x) := \Omega_\mu^3(x) .
  \label{dWl}
\end{eqnarray}
Now,  the difference
between the abelian (diagonal) Wilson loop (\ref{dWl}) and the full
non-Abelian Wilson loop (\ref{fWl}) has become clear. 
The diagonal Wilson loop is obtained from the full non-Abelian Wilson
loop by neglecting the $n^A(x) {\cal A}_\mu^A(x)$ part and the
functional integral 
$\int [d\mu({\bf n})]$ along the loop $C$.
Therefore, the deviation of the diagonal Wilson loop from the full
Wilson loop can be determined by estimating the effect from 
$n^A(x) {\cal V}_\mu^A(x)$.
\par
If the gauge field ${\cal A}_\mu(x)$ is not singular, the first piece
$U(x) {\cal A}_\mu(x) U^\dagger(x)$ of ${\cal A}_\mu^U(x)$  is
non-singular and does not give rise to magnetic current.
On the contrary, the second piece $\Omega_\mu(x)$ 
does give the non-vanishing magnetic monopole current 
(see e.g. \cite{KondoI}).
 According to Monte Carlo simulation on the
lattice
\cite{review}, the magnetic monopole part gives the most
dominant contribution in various quantities characterizing the
low-energy physics of QCD, e.g., string tension, chiral condensate,
topological charge, etc.
This phenomenon is called the magnetic monopole dominance.
\par
Therefore, it is expected that the most important degrees of
freedom for the low-energy physics comes from the second piece
$\Omega_\mu(x)$ of ${\cal A}_\mu^U(x)$ rather than the first piece,
$U(x) {\cal A}_\mu(x) U^\dagger(x)$.
Therefore, we have decomposed the YM theory into two
parts, i.e., the contribution from  the part
$\Omega_\mu(x)$ and the remaining part in section 2.

\par
Standing on this viewpoint, we recall the calculation of the abelian
(diagonal) field strength in four-dimensional YM theory.
The identity \cite{KondoI} for $\Omega_\mu$,
\begin{eqnarray}
 \partial_\mu \Omega_\nu(x)  - \partial_\nu \Omega_\mu(x)
 = ig[\Omega_\mu(x), \Omega_\nu(x)] + {i \over g}
U(x)[\partial_\mu, \partial_\nu]U^\dagger(x) .
\label{ident}
\end{eqnarray}
leads to
\begin{eqnarray}
 f_{\mu\nu}^\Omega(x) := 
 \partial_\mu \Omega_\nu^3(x)  - \partial_\nu \Omega_\mu^3(x)
 =   C_{\mu\nu}^{[\Omega]}(x) + {i \over g}
(U(x)[\partial_\mu,
\partial_\nu]U^\dagger(x))^{(3)} ,
\label{dec}
\end{eqnarray}
where   
\begin{eqnarray}
 C_{\mu\nu}^{[\Omega]} &:=& (ig[\Omega_\mu, \Omega_\nu])^{(3)}
 = g \epsilon^{ab3} \Omega_\mu^a \Omega_\nu^b 
 = ig (\Omega_\mu^+ \Omega_\nu^- - \Omega_\mu^- \Omega_\nu^+)  
\\ 
 &=& {1 \over g} 
 \sin \theta (\partial_\mu \theta \partial_\nu \varphi
 - \partial_\mu \varphi \partial_\nu \theta) .
\end{eqnarray}
Note that $C_{\mu\nu}^{[\Omega]}$ is generated from the off-diagonal
gluon fields, $\Omega_\mu^1, \Omega_\mu^2$.
\par
We can identity the first and second parts of right-hand-side (RHS)
of (\ref{dec})  with the  the magnetic monopole and the Dirac
string contributions respectively. 
This is clearly seen by writing (\ref{dec}) explicitly using Euler
angles,
\begin{eqnarray}
 f_{\mu\nu}^\Omega 
 = - {1 \over g}\sin \theta (\partial_\mu \theta \partial_\nu \varphi
 - \partial_\mu \varphi \partial_\nu \theta)
 + {1 \over g}([\partial_\mu, \partial_\nu] \chi 
 + \cos \theta  [\partial_\mu, \partial_\nu] \varphi) .
\end{eqnarray}
The magnetic charge is given by
\begin{eqnarray}
g_m(V^{(3)}) &=&  
{1 \over 2g}\int_{S^{(2)}} d^2
\sigma_{\rho\sigma} \epsilon_{\mu\nu\rho\sigma} 
 \sin \theta (\partial_\mu \theta \partial_\nu \varphi
 - \partial_\mu \varphi \partial_\nu \theta) .
\label{magc1}
\end{eqnarray}
The magnetic charge (\ref{magc1}) is  quantized
\cite{KondoI}, since the integrand is the Jacobian
from
$S^2$ to $S^2$ and
\begin{eqnarray}
 \Pi_2(SU(2)/U(1)) = \Pi_2(S^2) = Z .
\end{eqnarray}
Then the magnetic charge $g_m$ satisfies the
Dirac quantization condition,
\begin{eqnarray}
 g_m ={2\pi n \over g} , \quad {\rm or} \quad g g_m = 2\pi n \  (n
\in Z) .
 \label{DQ}
\end{eqnarray}
We can give the second definition of the magnetic charge $g_{m}$ as
the contribution from the Dirac string,
\begin{eqnarray}
 g_{DS}(V^{(3)}) &=&  
{1 \over 2 g}\int_{S^{(2)}} d^2
\sigma_{\rho\sigma} \epsilon_{\mu\nu\rho\sigma} 
 ([\partial_\mu, \partial_\nu] \chi 
 + \cos \theta  [\partial_\mu, \partial_\nu] \varphi) .
 \label{defmc}
\end{eqnarray}
If we choose
$\chi=-\varphi$  (resp. $\chi=+\varphi$) using residual U(1) gauge
invariance, then the Dirac string appears on the negative
(resp. positive)
$Z$ axis, i.e., $\theta=\pi$ (resp. $\theta=0$).  In this case,  the
surface integral (\ref{defmc}) reduces to the line integral around
the string,
\begin{eqnarray}
 g_{DS}(V^{(3)}) =  
{1 \over 2 g} \int_{S^{(2)}} d  \sigma_{\mu\nu} 
\epsilon_{\mu\nu\rho\sigma}
 [\partial_\rho, \partial_\sigma] \varphi(x) 
=
- {1 \over 2 g} \int_{S^{(1)}} d  \sigma_{\mu\nu\rho} 
\epsilon_{\mu\nu\rho\sigma} \partial_\rho \varphi(x) .
\end{eqnarray}
This gives the same result (\ref{magc1}) but with the minus sign,
which is in consistent with 
\begin{eqnarray}
  \Pi_1(U(1)) = Z .
\end{eqnarray}
Actually, two descriptions  (\ref{magc1}) and (\ref{defmc}) are
equivalent and the above argument can be generalized to more general
gauge group, as suggested from
\begin{eqnarray}
 \Pi_2(SU(N)/U(1)^{N-1}) = \Pi_1(U(1)^{N-1}) = Z^{N-1} .
\end{eqnarray}

\par
Eq.~(\ref{ident}) implies
\begin{eqnarray}
 {\cal F}_{\mu\nu}^U(x) 
 := \partial_\mu \Omega_\nu(x)  - \partial_\nu \Omega_\mu(x)
 - ig[\Omega_\mu(x), \Omega_\nu(x)]
 \equiv {i \over g}
 U(x) [\partial_\mu, \partial_\nu ]U^\dagger(x) ,
\label{fst}
\end{eqnarray}
if  the contribution from 
$U(x) {\cal A}_\mu(x) U^\dagger(x) $ 
is completely neglected. 
Here the RHS is identified with the contribution from the Dirac
string. 
The existence of Dirac string in the RHS of (\ref{fst}) reflects the
fact that the field strength ${\cal F}_{\mu\nu}^U(x)$ does contain
the magnetic monopole contribution.  
Thus we have obtained a gauge theory
with magnetic monopoles starting from the YM theory (without any
scalar field).  Therefore, MAG enables us to deduce the magnetic
monopole without introducing the scalar field, in contrast to the 't
Hooft-Polyakov monopole \cite{tHooft74,Polyakov74}.

\subsection{Magnetic monopole and non-Abelian Stokes theorem}
\par
We show that the magnetic monopole does appear in the gauge
invariant Wilson loop of YM theory through the NAST.
The NAST gives a gauge-invariant description of the magnetic
monopole in YM theory.
\par
The second term in the exponent
(\ref{NAST}) can be rewritten as a surface integral inside the closed
contour of the Wilson loop. The parameterization of SU(2) matrix $U$
by the Euler angles leads to
\begin{eqnarray}
 \Omega_\mu^3(x) 
:=  {\rm tr} \left(\sigma_3 {i \over g} U(x) \partial_\mu
U^\dagger(x) \right)
  = {1 \over g}[\partial_\mu \chi(x) + \cos \theta(x) \partial_\mu
\varphi(x)] .
\end{eqnarray}
Then the second term in the exponent of (\ref{NAST}) reads 
\begin{eqnarray}
iJ \oint_C dx^\mu \Omega_\mu^3(x) 
&=&  iJ \oint_C dx^\mu {\rm tr} \left(\sigma_3 {i \over g} U(x)
\partial_\mu U^\dagger(x) \right)
\nonumber\\
&=& i{J \over g} \oint_C dx^\mu  [\partial_\mu \chi(x) + \cos
\theta(x)
\partial_\mu \varphi(x) ] .
\end{eqnarray}
This is rewritten as a surface integral using the standard
Abelian Stokes theorem,
\begin{eqnarray}
iJ \oint_C dx^\mu \ \Omega_\mu^3(x) 
&=& iJ \int_S d^2z \ \epsilon_{\mu\nu} 
(\partial_\mu \Omega_\nu^3 - \partial_\nu \Omega_\mu^3) 
= iJ \int_S d^2z \ \epsilon_{\mu\nu}  f_{\mu\nu}^\Omega .
\end{eqnarray}
By making use of a unit vector ${\bf n}$, this is further rewritten
as
\cite{KondoII}
\begin{eqnarray}
iJ \oint_C dx^\mu  \Omega_\mu^3(x) 
&=& i{J \over g} \int_S d^2z \  \epsilon^{ABC} \epsilon_{\mu\nu} n^A
\partial_\mu n^B \partial_\nu n^C
\nonumber\\
&=&  i{J \over g} \int_S d^2z  \ \epsilon_{\mu\nu} 
{\bf n} \cdot (\partial_\mu {\bf n} \times \partial_\nu {\bf n}) 
\nonumber\\
&=&   i{8\pi J \over g}  Q_S ,
\end{eqnarray}
where  $Q$ is the topological
charge of the ${\bf n}$ field  \cite{KondoII} in the area $S$,
\begin{eqnarray}
  Q_S := {1 \over 8\pi} \int_S d^2z \ \epsilon_{\mu\nu} 
{\bf n} \cdot (\partial_\mu {\bf n} \times \partial_\nu {\bf n}) 
= {1 \over 4\pi} \int_S d^2 \sigma_{\mu\nu} \  
{\bf n} \cdot (\partial_\mu {\bf n} \times \partial_\nu {\bf n}) .
\end{eqnarray}
On the other hand, the first term in the exponent (\ref{NAST}) is
rewritten as
\begin{eqnarray}
  i J \oint dx^\mu \ {\rm tr}
 \left\{  \sigma_3 \left[ 
  U {\cal V}_\mu(x)U^\dagger \right] \right\}
  = i J \oint dx^\mu \  
   {\cal V}_\mu^A(x) n^A(x) 
   =  i J \oint dt \  
 {\cal V}^A(t) n^A(t) .
\end{eqnarray}
Thus we obtain another version of NAST,
\begin{eqnarray}
 && W^C [{\cal A}]  
 \nonumber\\ 
 &=& 
 \int [d\mu({\bf n})] 
 \exp \left\{ i J \left[ \oint_C dx^\mu \  
   n^A(x){\cal V}_\mu^A(x)  
  +  {1 \over g} \int_S d^2\sigma_{\mu\nu} 
{\bf n} \cdot (\partial_\mu {\bf n} \times \partial_\nu {\bf n}) 
\right] \right\} 
 \nonumber\\ 
 &=&  \int  [d\mu({\bf n})] 
 \exp \left\{  i J \left[ \oint_C dt \  
 n^A(t){\cal V}^A(t)  
 + {1 \over g} \int_S d^2\sigma_{\mu\nu} 
{\bf n} \cdot (\partial_\mu {\bf n} \times \partial_\nu {\bf n}) 
\right] \right\} .
 \label{Wlf2}
\end{eqnarray}
\par
Furthermore, the first term in the exponent is  
rewritten as
\begin{eqnarray}
 i J \oint_C dx^\mu \  
   n^A(x){\cal V}_\mu^A(x)  
   =   i J \int_S d^2\sigma^{\mu\nu} {1 \over 2}
   [\partial_\mu( n^A(x){\cal V}_\nu^A(x)) 
   - \partial_\nu( n^A(x){\cal V}_\mu^A(x))] .
\end{eqnarray}
Therefore, a manifestly gauge-invariant formula of the non-Abelian
Wilson loop has been obtained
\cite{DP89,DP96}
\begin{eqnarray}
  W^C [{\cal A}]  
= 
 \int [d\mu({\bf n})] 
 \exp \left\{ i {J \over 2} \int_S d^2\sigma^{\mu\nu}  \
G_{\mu\nu}(x)
\right\} ,
 \label{Wlf3}
\end{eqnarray}
with the gauge-invariant tensor field \cite{AFG75},
\begin{eqnarray}
 G_{\mu\nu}(x) :=  \partial_\mu(n^A(x){\cal V}_\nu^A(x)) 
   - \partial_\nu(n^A(x){\cal V}_\mu^A(x))
   - {1 \over g}{\bf n}(x) \cdot (\partial_\mu {\bf n}(x) \times
\partial_\nu {\bf n}(x)).
\label{tH1}
\end{eqnarray}
This is nothing but the 'tHooft tensor
\cite{tHooft74,Polyakov74,AFG75} if we identify $n^A$ with the
direction of the elementary Higgs field,
\begin{eqnarray}
 \hat \phi^A :=\phi^A/|\phi|, \quad |\phi|:= \sqrt{\phi^A \phi^A} . 
\end{eqnarray}
\par
The tensor (\ref{tH1})
gives a SU(2) gauge-invariant definition for the electromagnetic
field tensor, since 
using the covariant derivative,
\begin{eqnarray}
 D_\mu^{AB} :=  \partial_\mu \delta^{AB} - g \epsilon^{ABC} {\cal
A}_\mu^C ,
\end{eqnarray}
it is rewritten as
\begin{eqnarray}
 G_{\mu\nu}(x) &:=&   n^A(x) {\cal F}_{\mu\nu}^A(x)
   - {1 \over g}\epsilon^{ABC}{\bf n}^A(x) (D_\mu {\bf n}(x))^B
   (D_\nu {\bf n}(x))^C
\\
&=& {\rm tr}\left[{\bf n}(x) {\cal F}_{\mu\nu}(x)
   - {1 \over g}{\bf n}(x) (D_\mu {\bf n}(x))(D_\nu {\bf n}(x))
\right], 
\label{tH2}
\end{eqnarray}
where we have used
\begin{eqnarray}
 [\sigma^A , \sigma^B ] = 2i \epsilon^{ABC} \sigma^C, \quad
 {\rm tr}(\sigma^A \sigma^B) = 2\delta^{AB}, \quad
 {\rm tr}(\sigma^A \sigma^B \sigma^C) = 2i \epsilon^{ABC} .
\end{eqnarray}
Note that both terms in (\ref{tH2}) are gauge invariant, because
under the gauge transformation 
${\bf n}(x), D_\mu {\bf n}(x)$ and ${\cal F}_{\mu\nu}(x)$ transform
as the adjoint representation,
\begin{eqnarray}
  {\bf n}(x) &\rightarrow& U(x) {\bf n}(x) U^\dagger(x) ,
  \nonumber\\
  D_\mu {\bf n}(x) &\rightarrow& U(x) D_\mu {\bf n}(x) U^\dagger(x) ,
  \nonumber\\
   {\cal F}_{\mu\nu}(x) &\rightarrow& U(x) {\cal F}_{\mu\nu}(x)
U^\dagger(x)  .
\end{eqnarray}
\par
The Wilson loop is the evolution operator for spin $J{\bf n}$ in a
time-dependent "external (magnetic) field" ${\cal V}_\mu(t)$ and the
Wess-Zumino term,
\begin{eqnarray}
 S_{WZ} := \int d^2\sigma_{\mu\nu} \
{\bf n} \cdot (\partial_\mu {\bf n} \times \partial_\nu {\bf n})
\end{eqnarray}
fixes the representation to which the spin belongs.
The non-Abelian Wilson loop measures the flux of magnetic monopole
through the area $S$ enclosed by the Wilson loop $C$ where the
magnetic monopole is generated from the topological non-trivial
configuration of ${\bf n}(x)$.
\par
Unlike the usual electromagnetic field tensor, the tensor (\ref{tH1})
  has a dual with non-zero divergence, i.e., non-vanishing magnetic
monopole current,
\begin{eqnarray}
 k_\mu = {1 \over 2} \epsilon_{\mu\nu\rho\sigma} \partial^\nu
G^{\rho\sigma} 
=  {1 \over 2g} \epsilon_{\mu\nu\rho\sigma} \partial^\nu
{\bf n} \cdot (\partial_\rho {\bf n} \times \partial_\sigma {\bf n})
=  {1 \over 2g} \epsilon_{\mu\nu\rho\sigma} \partial^\nu
[{\bf n} \cdot (\partial_\rho {\bf n} \times \partial_\sigma {\bf
n})] .
\end{eqnarray}
The monopole current $k_\mu$ is a conserved topological current,
$\partial^\mu k_\mu \equiv 0$.
Although the $k_\mu$ is written as a total divergence, it can give
non-vanishing magnetic charge (\ref{magc1}),
\begin{eqnarray}
 g_m = \int_{V^{(3)}} d^3 x \ k_0 = {2\pi n \over g}.
\end{eqnarray}
In the region where 
${\bf n}=(0,0,1)$, the 't Hooft tensor reads
\begin{eqnarray}
 G_{\mu\nu}(x) =  \partial_\mu {\cal V}_\nu^3(x) 
   - \partial_\nu {\cal V}_\mu^3(x)  ,
\end{eqnarray}
and the magnetic current vanishes identically, $k_\mu
\equiv 0$.

\subsection{Abelian magnetic monopole dominance}
\par
Note that we can replace ${\cal A}_\mu$ (appearing in the argument
of the exponent in the NAST (\ref{NAST})) with ${\cal V}_\mu$ which
has been defined in the reformulation of the YM theory.  This is
shown as follows. If
${\cal A}_\mu (x)$ is the gauge rotation of
${\cal V}_\mu(x)$ by
$\tilde U(x)$,   
\begin{eqnarray}
{\cal A}_\mu (x) &:=& 
 \tilde U(x) {\cal V}_\mu(x) \tilde U^\dagger(x) + {i \over g}
\tilde U(x) \partial_\mu \tilde U^\dagger(x) 
= {\cal V}_\mu^{\tilde U}(x) , 
\end{eqnarray}
then 
\begin{eqnarray}
 {\cal A}_\mu^U (x) &:=& U(x) {\cal A}_\mu(x)U^\dagger(x) 
  + {i \over g} U(x) \partial_\mu U^\dagger(x) 
  \nonumber\\
 &=& 
  (U(x)\tilde U(x)) {\cal V}_\mu(x) (U(x)\tilde U(x))^\dagger 
  + {i \over g} (U(x)\tilde U(x)) \partial_\mu (U(x)\tilde
U(x))^\dagger 
  \nonumber\\
 &=& {\cal V}_\mu^{U\tilde U}(x).
\end{eqnarray}
As the new matrix $U\tilde U$ is also an element of $G$,  we can
absorb this change into the invariant Haar measure
$[d\mu_C({\bf n})]$.   Therefore we
can write the NAST (\ref{NAST2}) as
\begin{eqnarray}
 && W^C [{\cal A}] :=  {\rm tr} \left[ {\cal P} 
 \exp \left( i \oint_C {\cal A}_\mu^A(x) T^A dx^\mu
 \right) \right] 
 \nonumber\\ 
 &=& 
 \int  [d\mu_C({\bf n})]  
 \exp \left( i J \oint_C dx^\mu \ {\rm tr}
 \left\{  \sigma_3 \left[ 
  U {\cal V}_\mu(x)U^\dagger 
  + {i \over g} U \partial_\mu U^\dagger  \right]
 \right\} \right) 
 \nonumber\\ 
 &=&  \int  [d\mu_C({\bf n})]  
 \exp \left( i J \oint dt \ {\rm tr} \left\{ \sigma_3
\left[  U {\cal V}(t) U^\dagger 
 + {i \over g} U {d \over dt} U^\dagger \right] \right\}  \right) ,
 \label{NAST2}
\end{eqnarray}
and the expectation value of the Wilson loop reads
\begin{eqnarray}
&& \langle W^C[{\cal A}] \rangle_{YM} 
 \nonumber\\
 &=& \int d\mu_C({\bf n}) \
  \Biggr\langle \exp  i J \left[ \oint_C dx^\mu \  
   n^A(x){\cal V}_\mu^A(x)  
  +  {1 \over g} \int_S d^2z \ \epsilon_{\mu\nu} 
{\bf n} \cdot (\partial_\mu {\bf n} \times \partial_\nu {\bf n}) 
\right]   \Biggr\rangle ,
\end{eqnarray}
where the expectation value is written according to (\ref{deform}) as
\begin{eqnarray}
 && \left\langle \exp \left\{ i J \left[ \oint_C dx^\mu \  
   n^A(x){\cal V}_\mu^A(x)  
  +  {1 \over g} \int_S d^2z \ \epsilon_{\mu\nu}
{\bf n} \cdot (\partial_\mu {\bf n} \times \partial_\nu {\bf n}) 
\right] \right\}  \right\rangle_{YM} 
\nonumber\\&&
= Z_{YM}^{-1} \int [dU][d{\cal C}][d\bar {\cal C}]
 [d\phi]
e^{i S_{TFT}[\Omega_\mu, {\cal C}, \bar {\cal C}, \phi]}
  e^{ i{J \over g} \int_S d^2z \ \epsilon_{\mu\nu}
{\bf n} \cdot (\partial_\mu {\bf n} \times \partial_\nu {\bf n}) }
 \nonumber\\&&   
 \times
\int [d{\cal V}_\mu][d\gamma][d\bar \gamma][d\beta]
 e^{ i S_{pYM}[{\cal V},\gamma, \bar \gamma, \beta]}
\nonumber\\&&  
\times
 e^{i \int d^Dx \left(  i{\cal V}_\mu^A {\cal M}_\mu^A[U]
 + {i \over 2} {\cal V}_\mu^A {\cal V}_\mu^B {\cal K}^{AB}[U]
\right) }
  e^{i J  \oint_C dx^\mu    n^A(x){\cal V}_\mu^A(x)  } .
\end{eqnarray}
The denominator, i.e. the partition function $Z_{YM}$ is equal to
\begin{eqnarray}
 Z_{YM} = \left\langle  \left\langle 
 e^{i \int d^Dx \left(  i{\cal V}_\mu^A {\cal M}_\mu^A[U]
 + {i \over 2} {\cal V}_\mu^A {\cal V}_\mu^B {\cal K}^{AB}[U]
\right) }  
\right\rangle_{pYM} Z_{pYM} \right\rangle_{TFT} Z_{TFT},
\label{deno}
\end{eqnarray}
where $Z_{pYM}$ is the partition function of perturbative sector of
the YM theory,
\begin{eqnarray}
 Z_{pYM} :=
\int [d{\cal V}_\mu][d\gamma][d\bar \gamma][d\beta]
 e^{ i S_{pYM}[{\cal V},\gamma, \bar \gamma, \beta]} ,
\end{eqnarray}
and $Z_{TFT}$ is the partition function of the TFT,
\begin{eqnarray}
 Z_{TFT} :=
\int [dU][d{\cal C}][d\bar {\cal C}]
 [d\phi]
e^{i S_{TFT}[\Omega_\mu, {\cal C}, \bar {\cal C}, \phi]} .
\end{eqnarray}
The numerator is equal to
\begin{eqnarray}
 && \Biggr\langle \left\langle 
 e^{i \int d^Dx \left(  i{\cal V}_\mu^A {\cal M}_\mu^A[U]
 + {i \over 2} {\cal V}_\mu^A {\cal V}_\mu^B {\cal K}^{AB}[U]
\right) } e^{ i J  \oint_C dx^\mu  n^A(x){\cal V}_\mu^A(x)  }
\right\rangle_{pYM}  Z_{pYM}
\nonumber\\&&  \quad \quad
 \times e^{i {J \over g} \int_S d^2z \ \epsilon_{\mu\nu}
{\bf n} \cdot (\partial_\mu {\bf n} \times \partial_\nu {\bf n}) }
\Biggr\rangle_{TFT} Z_{TFT} .
\label{nume}
\end{eqnarray}
\par
The expectation value  of the Wilson loop is given by the ratio, 
(\ref{nume})/(\ref{deno}).
In (\ref{deno}), the argument of the exponential  including 
${\cal M}_\mu^A[U]$ and ${\cal K}^{AB}[U]$ (\ref{defMK}) is written
in the BRST exact form,
\begin{eqnarray}
 e^{i\int d^Dx \left(  i{\cal V}_\mu^A {\cal M}_\mu^A[U]
 + {i \over 2} {\cal V}_\mu^A {\cal V}_\mu^B {\cal K}^{AB}[U]
\right)}  = e^{i\{ Q_B, * \} } .
\end{eqnarray}
Expanding this exponential and using
the fact that  
\begin{eqnarray}
 Q_B^\dagger = Q_B,  \quad
 Q_B |0\rangle_{TFT}=0, \quad Q_B^2=0 ,
\end{eqnarray}
we see that the partition function in the {\it absence of external
sources} has the decomposition,
\begin{eqnarray}
 Z_{YM} =   Z_{pYM} Z_{TFT} .
\label{deno2}
\end{eqnarray}
Thus  the expectation value of the Wilson loop  is written as
\begin{eqnarray}
  \langle W^C[{\cal A}] \rangle_{YM} &=& 
 \Biggr\langle \left\langle 
 e^{i \int d^Dx \left(  i{\cal V}_\mu^A {\cal M}_\mu^A[U]
 + {i \over 2} {\cal V}_\mu^A {\cal V}_\mu^B {\cal K}^{AB}[U]
\right) } e^{ i J  \oint_C dx^\mu  n^A(x){\cal V}_\mu^A(x)  }
\right\rangle_{pYM}   
\nonumber\\&&  \quad \quad \quad \quad \quad \quad
 \times e^{i {J \over g} \int_S d^2z \ \epsilon_{\mu\nu}
{\bf n} \cdot (\partial_\mu {\bf n} \times \partial_\nu {\bf n}) }
\Biggr\rangle_{TFT} .
\label{W1}
\end{eqnarray}
By repeating similar arguments, the Wilson loop is
 cast into the form,
\begin{eqnarray}
  \langle W^C[{\cal A}] \rangle_{YM} = 
 \left\langle 
\left\langle 
 e^{i J  \oint_C dx^\mu  n^A(x){\cal V}_\mu^A(x)  }
\right\rangle_{pYM} 
e^{i {J \over g} \int_S d^2z \ \epsilon_{\mu\nu} {\bf n} \cdot
(\partial_\mu {\bf n} \times \partial_\nu {\bf n}) }
 \right\rangle_{TFT}  .
\label{W2}
\end{eqnarray}
\par
The perturbative part is expanded into
\begin{eqnarray}
&& \left\langle 
 e^{i J  \oint_C dx^\mu  n^A(x){\cal V}_\mu^A(x)  }
\right\rangle_{pYM}  
\nonumber\\&& 
= 1 - {1 \over 2} J^2 \oint_C dx^\mu \oint_C dy^\nu n^A(x) n^B(y)
\left\langle {\cal V}_\mu^A(x) {\cal V}_\nu^B(y) \right\rangle_{pYM}
+ O(g^4) ,
\label{W3}
\end{eqnarray}
where we have used 
$
 \left\langle {\cal V}_\mu^A(x)  \right\rangle_{pYM}= 0 .
$
Then we can write
\begin{eqnarray}
 \langle W^C[{\cal A}] \rangle
   &=& 
 \left\langle  e^{i {J \over g} \int_S d^2z \ \epsilon_{\mu\nu}
{\bf n} \cdot (\partial_\mu {\bf n} \times \partial_\nu {\bf n}) }
 \right\rangle_{TFT} 
 \nonumber\\&&
\times \Biggr[ 1 -  
 {1 \over 2} J^2 \oint_C dx^\mu \oint_C dy^\nu 
\left\langle {\cal V}_\mu^A(x) {\cal V}_\nu^B(y) \right\rangle_{pYM}
\nonumber\\&&  \quad \quad \quad 
\times
 {\left\langle  n^A(x) n^B(y) 
 e^{i {J \over g} \int_S d^2z \ \epsilon_{\mu\nu}
{\bf n} \cdot (\partial_\mu {\bf n} \times \partial_\nu {\bf n}) }
\right\rangle_{TFT}  
\over 
\left\langle   e^{i {J \over g} \int_S d^2z \ \epsilon_{\mu\nu}
{\bf n} \cdot (\partial_\mu {\bf n} \times \partial_\nu {\bf n}) }
\right\rangle_{TFT}
}
+ O(g^4) \Biggr] .
\label{W4}
\end{eqnarray}
\par
Owing to the dimensional reduction, the expectation value of the
diagonal Wilson loop 
$
 \left\langle  e^{i {J \over g} \int_S d^2z \ \epsilon_{\mu\nu}
{\bf n} \cdot (\partial_\mu {\bf n} \times \partial_\nu {\bf n}) }
 \right\rangle_{TFT} 
$
in the RHS of (\ref{W4}) 
in the four-dimensional TFT (\ref{TFT})
 is reduced to that in the two-dimensional NLSM (\ref{NLSM}),
when $C$ is  planar,
\begin{eqnarray}
 \langle W^C[a^\Omega] \rangle_{TFT_4}
 =  \left\langle  e^{i {J \over g} \int_S d^2z \ \epsilon_{\mu\nu} 
{\bf n} \cdot (\partial_\mu {\bf n} \times \partial_\nu {\bf n}) }
 \right\rangle_{TFT_4} 
 = \left\langle  e^{i {J \over g} \int_S d^2z \ \epsilon_{\mu\nu}
{\bf n} \cdot (\partial_\mu {\bf n} \times \partial_\nu {\bf n}) }
 \right\rangle_{NLSM_2}  .
\end{eqnarray}
The quantity $Q_S[{\bf n}]$ defined by
\begin{eqnarray}
Q_S[{\bf n}] := { 1 \over 8\pi } \int_S  d^2z \ \epsilon_{\mu\nu}  
{\bf n} \cdot (\partial_\mu {\bf n} \times \partial_\nu {\bf n})
=   n^{in}_+ - n^{in}_-  , 
\end{eqnarray}
is an integer and counts the instanton--anti-instanton charge 
($n^{in}_+ - n^{in}_-$)
inside
the Wilson loop. By summing up the instanton and anti-instanton
contributions in the two-dimensional NLSM, we have obtained the area
law for the diagonal Wilson loop in the previous article
\cite{KondoII},
\begin{eqnarray}
 \langle W^C[a^\Omega] \rangle_{TFT_4}
= 
 \left\langle  e^{i {J \over g} 8\pi Q_S[{\bf n}] }
 \right\rangle_{NLSM_2}  
\cong  e^{-\sigma_{Abel} A(C)},  
\end{eqnarray}
where $A(C)$ is the area enclosed by the Wilson loop $C$.
We call the coefficient in the area decay
$\sigma_{Abel}$ the Abelian string tension.
The naive instanton calculus based on the dilute
instanton gas approximation \cite{KondoII} leads to
\begin{eqnarray}
 \sigma_{Abel} =   2B e^{-S_1} \left[ 1 - \cos 
\left( {2\pi q \over g}\right) \right]  ,
\quad S_1= {4\pi^2 \over g^2} ,
\end{eqnarray}
where $B$ is a constant with the mass-squared dimension, 
$B \sim m_A^2$ and
$S_1= 4\pi^2/g^2$ is the action for one instanton.  
Here we have neglected to write the perimeter decay part which can
be generated by instantons and anti-instantons located just on the
perimeter of the Wilson loop.
\par
Now we proceed to estimate the remaining terms.
To simplify the perturbation calculation  in the RHS of  (\ref{W4}),
we take the Feynman gauge in the perturbative sector where
the propagator reads 
\begin{eqnarray}
\left\langle {\cal V}_\mu^A(x) {\cal V}_\nu^B(y) \right\rangle_{pYM}
&=& \delta^{AB} \delta_{\mu\nu} G(x,y) ,
\\
G(x,y) &=& \int {d^4 p \over (2\pi)^4} 
e^{ip(x-y)}{g^2 \over p^2}
 = {g^2 \over 4\pi^2} {1 \over |x-y|^2} .
\end{eqnarray}
Then we obtain
\begin{eqnarray}
  \langle W^C[{\cal A}] \rangle_{YM} &=&   
 \left\langle  e^{i {J \over g} 8\pi Q_S[{\bf n}]}
 \right\rangle_{NLSM_2} 
  \Biggr[ 1 -   
 {1 \over 2} J^2 \oint_C dx^\mu \oint_C dy^\mu G(x,y) 
\nonumber\\&&  \quad \quad \quad 
\times
 {\left\langle  {\bf n}(x) \cdot {\bf n}(y) 
 e^{i {J \over g} 8\pi Q_S[{\bf n}]}
\right\rangle_{NLSM_2}  
\over 
\left\langle   e^{i {J \over g} 8\pi Q_S[{\bf n}]}
\right\rangle_{NLSM_2}
}
+ O(g^4) \Biggr] .
\label{W5}
\end{eqnarray}
where we have used the dimensional reduction  \cite{KondoII} for the
correlation function,
\begin{eqnarray}
 && \left\langle  {\bf n}(x) \cdot {\bf n}(y) 
 e^{i {J \over g} 8\pi Q_S[{\bf n}]}
\right\rangle_{TFT_4}  
= \left\langle  {\bf n}(x) \cdot {\bf n}(y) 
 e^{i {J \over g} 8\pi Q_S[{\bf n}]}
\right\rangle_{NLSM_2},
\nonumber\\
&& x, y \in C=\partial S \subset R^2 ,
\label{nn}
\end{eqnarray}
for the planar Wilson loop $C$, 

\par
A naive estimate for the expectation value in NLSM$_2$ is given by
considering the instanton contribution.  
The instanton solution in NLSM
is given by the field configuration ${\bf n}$ such that ${\bf n}$
approaches the same value ${\bf n}^{(0)}$ at infinity (see
\cite{KondoII})
\begin{eqnarray}
  {\bf n}(x) \rightarrow {\bf n}^{(0)} \quad (|x| \rightarrow
\infty) ,
\end{eqnarray}
where ${\bf n}^{(0)}$ is any unit vector,
${\bf n}^{(0)} \cdot {\bf n}^{(0)} = 1$.
Therefore, for large Wilson loop $C$
\begin{eqnarray}
 {\bf n}(x) \cdot {\bf n}(y) \rightarrow 
 {\bf n}^{(0)} \cdot {\bf n}^{(0)} = 1, 
 \quad (x, y \in C=\partial S) .
 \label{esti}
\end{eqnarray}
Here the configuration ${\bf n}^{(0)}\equiv (0,0,1)$ corresponds to
the topological trivial case, 
 $Q=0$.
The precise estimate of (\ref{nn}) can be done using the large $N$
expansion for O(N) NLSM.  In fact, for not so large $|x-y|$, the
two-point correlation function behaves as
(see e.g. \cite{Polyakov87})
\begin{eqnarray}
\left\langle  {\bf n}(x) \cdot {\bf n}(y) 
\right\rangle_{NLSM_2} 
= \left[ 1 - {N-2 \over 2\pi} {1 \over \beta} \ln {|x-y| \over
\epsilon} \right]^{{N-1 \over N-2}} ,
\label{corre}
\end{eqnarray}
where $\epsilon$ is a short distance cutoff.
\par
It turns out that the contribution of the last term in
(\ref{W5}) gives the perimeter law correction to the area law.
For large $T \gg R \gg 1$, we have (see Appendix and
\cite{Kogut83,BBJ84,Fischler77,Hughes80,Polaykov79,DV80})
\begin{eqnarray}
  - {1 \over 2}  \oint_C dx^\mu \oint_C dy^\mu  G(x,y)
    \cong  
  - {g^2 \over 2\pi^2} {T+R \over \epsilon} 
   + {g^2 \over 4\pi} {T \over R}    
   + {g^2 \over 2\pi^2} \ln  {R \over \epsilon}   .
   \label{peri}
\end{eqnarray}
It should be remarked that the perimeter decay in (\ref{peri}) comes
from the contribution of the coincident point,
$x=y$, (after regularization, $|x-y| \cong \epsilon \ll 1$, see
Appendix).
Similarly we can evaluate the higher-order terms which give the
running coupling constant in consistent with the asymptotic freedom.
These contributions from the perturbative sector should be compared
with the conventional calculation based on the perturbative QCD
\cite{Kogut83}.
\par
Exponentiating the contributions from the power-series expansion
\cite{Kogut83}, we obtain for
$g$ small,
\begin{eqnarray}
  \langle W^C[{\cal A}] \rangle_{YM}
 &&\cong  
\left\langle   e^{i {J \over g} 8\pi Q } \right\rangle_{NLSM_2}  
e^{- C g^2 (R+T) + {g^2 \over 4\pi} {T \over R} + C'}
\nonumber\\&&
=  e^{-\sigma_{Abel} RT - C g^2 (R+T) 
+ {g^2 \over 4\pi} {T \over R} + C' } ,   
\label{areaW}
\end{eqnarray}
where $C$ and $C'$ are constants.
The full non-Abelian string tension is defined by
\begin{eqnarray}
  \sigma := - \lim_{A(C) \rightarrow \infty}
  {1 \over A(C)} \ln \langle W^C[{\cal A}] \rangle_{YM} ,
\end{eqnarray}
where $A(C)$ is the area enclosed by the Wilson loop.
For a rectangular loop with side lengths $R$ and $T$, $A(C)=RT$.
The above result shows that the Abelian (diagonal) Wilson loop obeys
the same area law as  the Non-Abelian Wilson loop and that the
area law of the full non-Abelian Wilson loop is deduced from the
magnetic monopole contribution for the diagonal Wilson loop,
$
\left\langle   e^{i {J \over g} 8\pi Q } \right\rangle_{TFT} .
$
This implies the monopole dominance in the string tension of QCD
under the MAG. The deviation of the Abelian string tension from the
full non-Abelian string tension is given by
\begin{eqnarray}
  \sigma - \sigma_{Abel} 
  &&= - \lim_{A(C) \rightarrow \infty}
  {1 \over A(C)} \ln \langle W^C[{\cal A}] \rangle_{YM}
  + \lim_{A(C) \rightarrow \infty} {1 \over A(C)} \ln  
  \left\langle  e^{i {J \over g} 8\pi Q }
 \right\rangle_{TFT}
 \nonumber\\&&
 = \lim_{R,T \rightarrow \infty} {C g^2 (R+T) 
 - {g^2 \over 4\pi} {T \over R} - C' \over RT} 
 = 0.
 \label{devi}
\end{eqnarray}
Hence the deviation of the  string tension comes from the finite size
effect of the Wilson loop. 
For sufficiently large Wilson loop, $\sigma > \sigma_{Abel}$
and the off-diagonal contribution to the string
tension vanishes as 
$R,T \rightarrow \infty$.
In the large Wilson loop limit $R,T \rightarrow \infty$, the Abelian
string tension coincides exactly with the full non-Abelian string
tension,
$\sigma = \sigma_{Abel}$.
Thus Abelian and monopole dominance for the string tension can be
proved under the MAG according to the formulation of the YM theory as
a deformation of the TFT.
It is rather straightforward to
extend the above strategy to the case,
$G=SU(N), N \ge 3$.
\par
It should be remarked that, if the massive decay of the correlation
function (\ref{corre}) for large separation 
$|x-y| \gg 1 \ (x,y \in C)$ is incorporated in the above evaluation,
the Coulomb part 
$
{g^2 \over 4\pi} {T \over R}
$
in (\ref{peri}) will be replaced by the Yukawa part,
$
{g^2 \over 4\pi} {T \over R} e^{-mR} ,
$
where $m$ is the mass of the ${\bf n}$ field of the NLSM.
However, this effect does not change the conclusion for the string
tension.  In fact, the perimeter part is generated from the
coincidence limit $|x-y| \ll 1$.
Furthermore, if we averaged 
${\bf n}(x) \cdot {\bf n}(y)$ over all possible configurations, we
would have obtained,
\begin{eqnarray}
  \int d\mu_C({\bf n}) \
  {\bf n}(x) \cdot {\bf n}(y)  = \delta^{(2)}(x-y) ,
\end{eqnarray}
from a fact that ${\bf n}(x)$ and ${\bf n}(y)$ are independent for
the measure 
$d\mu_C({\bf n})$.   Consequently, only the coincident
contribution survives in (\ref{W5}) , which leads to the perimeter
decay correction alone in (\ref{devi}) (without the Coulomb or Yukawa
part).
\par

Monte Carlo simulation of lattice gauge theory
supports the finite size effect as a deviation of the string
tension, as argued by Suganuma et al.
\cite{SITA98} using the computer-assisted analytical study.

\subsection{Abelian dominance}

According to the NAST, one must average over
all gauge transformation in the coset SU(2)/U(1).  
Abelian dominance is the statement that in the true quantum vacuum,
the contributions to the $n$ average is approximated by the Abelian
projection.  This replaces $n^A(x){\cal V}_\mu^A(x)$ with
$n^3(x){\cal V}_\mu^3(x)$. 
In our standpoint, the contribution to the area law and
non-vanishing string tension comes from the topological term,
i.e., the second term in the exponent (\ref{Wlf2}), because the
first term can only give the perturbative correction around
non-trivial topological sector.  Actually, the first term may give a
long-range Coulomb potential in the topological trivial sector $Q=0$.
Therefore, according to the reformulation of YM theory as a
deformation of MAG TFT, the dominance of topological non-trivial
term (the second term) is an immediate consequence of the
formulation. This implies the monopole dominance in the string
tension.
In addition, the Abelian dominance is an immediate consequence of the
APEGT together with the above considerations.

\par
Here it should be remarked that the Abelian (diagonal) Wilson loop
in the non-Abelian gauge theory is not the same as the Wilson loop in
the Abelian gauge theory. 
In the Abelian U(1) gauge theory, the Wilson loop is given by
\cite{KondoIII}
\begin{eqnarray}
 && W^C [a] :=  
 \exp \left( i q \oint_C a_\mu(x) dx^\mu
 \right)  
 \nonumber\\ 
 &=& 
 \exp \left( i  q \oint_C dx^\mu \   \left[ 
   v_\mu(x) + {i \over g} U \partial_\mu U^\dagger  \right] \right) ,
   \quad U(x) = e^{i\varphi(x)} \in U(1) .
\end{eqnarray}

\subsection{Gluon self-interactions in the perturbative sector}

In the new reformulation of gauge theory \cite{KondoII}, QCD has
been identified with a perturbative deformation of the TQFT.  In
this reformulation the non-perturbative dynamics of QCD is saturated
by the TQFT.    This identification will be meaningful at least in
the low energy physics (including the quark confinement) by the
following reasons.  In principle,  of course, additional
non-perturbative dynamics could possibly come from the
self-interaction among the gluon fields reflecting the non-Abelian
nature of the gauge group.  However, additional non-perturbative
contributions to quark confinement are expected to be rather small,
if any.  This is because the recent numerical simulations
\cite{SY90,review} of lattice gauge theory with the maximal Abelian
gauge fixing have confirmed the magnetic monopole dominance as well
as the Abelian dominance in low-energy physics of QCD for various
quantities including the string tension. 
\par
Another reason from the theoretical viewpoint is as follows.
As shown in \cite{KondoI}, we can integrate out the off-diagonal
gluon fields in QCD to obtain the low-energy effective gauge theory
of QCD, i.e. APEGT.  Note that the APEGT is the Abelian gauge
theory.   Hence, at this stage, we do not worry so seriously about
the remaining gluon self-interactions which are identified as the
perturbative deformation to the TQFT in the reformulation of QCD. 
Then the non-vanishing magnetic monopole current
$k_\mu(x)$ is generated from the diagonal Abelian part
$a_\mu^\Omega(x)$ according to \cite{KondoI}.  
In addition, the result of the previous article \cite{KondoII} shows
that the condensation of magnetic monopoles in the four-dimensional
QCD  is deduced from the instanton (or vortex) condensation in  the
two-dimensional NLSM  obtained from the dimensional reduction of the
four-dimensional TQFT. Therefore, the low-energy dynamics of QCD in
the MAG is considered to be described well by the TQFT or its
dimensional reduction, i.e., NLSM.   In the low-energy region where
the APEGT is meaningful,  therefore, quark confinement will follow
from  these considerations without much difficulties by combining
the results of the previous articles
\cite{KondoI,KondoII} with the result of this article.

\appendix
\section{Evaluation of Wilson integral}
\setcounter{equation}{0}

In order to evaluate the Wilson integral, we choose a rectangular
contour with side lengths $R$ and $T$.  Then we have
\begin{eqnarray}
 && \oint_C dx_\mu \oint_C dy_\mu {1 \over |x-y|^2}
  \nonumber\\&&
  =  -2 \int_0^T dt' \int_0^T dt'' {1 \over R^2+(t'-t'')^2}
  - 2 \int_0^R dr' \int_0^R dr'' {1 \over T^2+(r'-r'')^2}
  \nonumber\\&& \quad
  + 2 \int_0^T dt' \int_0^T dt'' {1 \over (t'-t'')^2}
  + 2 \int_0^R dr' \int_0^R dr'' {1 \over (r'-r'')^2} .
  \label{integral}
\end{eqnarray}
Note that $dx_\mu dy_\mu$ implies that only integrations between
parallel sides give a contribution, i.e., no contribution between
neighboring sides where $dx_\mu dy_\mu=0$.
In the line integrals in the first (resp. second) lines of
(\ref{integral}),
$x$ and
$y$ run over opposite (resp. same) sides of the rectangle.
The integral over the opposite side is 
\begin{eqnarray}
  \int_0^T dt' \int_0^T dt'' {1 \over R^2+(t'-t'')^2}
= {2T \over R} \arctan {T \over R} - \ln \left( 1 + {T^2 \over R^2}
\right) . 
\end{eqnarray}
On the other hand, the integrals over the same sides diverge.  So we
omit the integral around the singularity by introducing the
infinitesimal parameter
$\epsilon$,
\begin{eqnarray}
 \int_0^T dt' \int_0^T dt'' {1 \over (t'-t'')^2}
 = 2 \int_0^{T-\epsilon} dt' \int_{t'+\epsilon}^T dt'' {1 \over
(t'-t'')^2}
 =  2 {T-\epsilon \over \epsilon}
  + 2 \ln  {\epsilon \over T} .
\end{eqnarray}
Summing all terms yields
\begin{eqnarray}
 && \oint_C dx_\mu \oint_C dy_\mu {1 \over |x-y|^2}
  \nonumber\\&&
  =  -2 \Biggr[
  {2T \over R} \arctan {T \over R} 
  +  {2R \over T} \arctan {R \over T} 
  - \ln \left( 1 + {T^2 \over R^2} \right)
  - \ln \left( 1 + {R^2 \over T^2} \right)
  \nonumber\\&& \quad \quad
  - 2 {T+R-2\epsilon \over \epsilon}
  - 2 \ln  {\epsilon \over T} - 2 \ln  {\epsilon \over R}
  \Biggr] .
\end{eqnarray}
For simplicity, we consider the rectangle which is much larger in
temporal direction than the spatial direction, $T \gg R (\gg
\epsilon)$,
\begin{eqnarray}
 \oint_C dx_\mu \oint_C dy_\mu 
  {1 \over |x-y|^2}
  \cong  -2 \Biggr[ \pi {T \over R} + 2 \ln  {R \over \epsilon} 
  - 2 {T+R-2\epsilon \over \epsilon} 
  \Biggr] .
\end{eqnarray}
Then we obtain
\begin{eqnarray}
 - {1 \over 2} \oint_C dx_\mu \oint_C dy_\mu 
 {g^2 \over 4\pi^2} {1 \over |x-y|^2}
  \cong  
  - {g^2 \over 2\pi^2} {T+R \over \epsilon} 
   + {g^2 \over 4\pi} {T \over R}    
   + {g^2 \over 2\pi^2} \ln  {R \over \epsilon}   .
   \label{integ}
\end{eqnarray}
In RHS of (\ref{integ}) the first term exhibits the perimeter law. 
The second term corresponds to the Coulomb law.  The constant term
represents the self-energy of a regularized point charge.  This
should be subtracted during the course of renormalization.  
On the lattice, $\epsilon$ is replaced with the lattice spacing.
The last logarithmic divergent term does not occur for the path $C$
with continuous tangent \cite{DV80}.
\par
In the Abelian gauge theory, the static potential is given by
\begin{eqnarray}
 V(R) &:=&	 - \lim_{T \rightarrow \infty} {1 \over T} \ln 
 \langle W^C[v] \rangle
\nonumber\\ 
 &=&  - \lim_{T \rightarrow \infty} {1 \over T} \ln 
 \left\langle  \exp \left[ i   \oint_C dx^\mu   v_\mu(x) \right]
\right\rangle_{pU(1)}  
 \nonumber\\ 
 &=& - \lim_{T \rightarrow \infty} {1 \over T} \ln 
 \exp \left[ -{1 \over 2}  \oint_C dx^\mu \oint_C dy^\nu 
\left\langle v_\mu(x) v_\nu(y)
\right\rangle_{pU1)} \right]
 \nonumber\\ 
 &=&  - \lim_{T \rightarrow \infty} {1 \over T} \ln 
 \exp \left( - {1 \over 2} \oint_C dx_\mu \oint_C dy_\mu 
 {g^2 \over 4\pi^2} {1 \over |x-y|^2} \right) 
 \nonumber\\ 
  &\cong&  
   {g^2 \over 2\pi^2} {1 \over \epsilon} 
   - {g^2 \over 4\pi} {1 \over R}    ,
\end{eqnarray}
where the last term gives the Coulomb potential.
The above evaluation of the Wilson loop shows that the perimeter
law follows from the contribution,
$x=y$.
For more details, see 
\cite{Kogut83,BBJ84,Fischler77,Hughes80,Polaykov79,DV80}.

\section*{Acknowledgments}
I would like to thank Volodya Miransky for suggestions on revising
the article and Giovanni Prosperi for pointing out the misleading
expressions in the first version of this article.
I am also grateful to Mauro Zeni for sending many comments on the 
articles \cite{KondoII,KondoIII} and Dmitri Antonov for
informing me of the information on more non-Abelian Stokes theorems.
This work is supported in part by
the Grant-in-Aid for Scientific Research from the Ministry of
Education, Science and Culture.


\end{document}